\documentclass[10pt,twoside]{article}

\usepackage{amssymb} 
\usepackage{amsmath} 
\usepackage{amsfonts}
\usepackage[dvips]{epsfig}
\usepackage[T1]{fontenc}

\newcommand{\vb}{v}

\newcommand{\vecr}{ {\bf r}}

\newcommand{\Vsr}{V_{\scriptscriptstyle SR}}
\newcommand{\ai}{{\scriptscriptstyle \alpha}_i}
\newcommand{\aj}{{\scriptscriptstyle \alpha}_j}
\newcommand{\xid}{\xi_{\scriptscriptstyle D}}
\newcommand{\soma}{\sum_{ \alpha}}
\newcommand{\ea}{e_{\alpha}}
\newcommand{\ma}{m_{\alpha}}
\newcommand{\mg}{m_{\gamma}}
\newcommand{\ead}{e_{\alpha}^2}
\newcommand{\eap}{e_{\alpha'}}
\newcommand{\za}{z_{\alpha}}
\newcommand{\mua}{\mu_{\alpha}}
\newcommand{\laa}{\lambda_{\alpha}}
\newcommand{\laap}{\lambda_{\alpha'}}
\newcommand{\lag}{\lambda_{\gamma}}
\newcommand{\li}{\lambda_{{\alpha}_i}}
\newcommand{\lj}{\lambda_{{\alpha}_j}}
\newcommand{\ei}{e_{{\alpha}_i}}
\newcommand{\ej}{e_{{\alpha}_j}}
\newcommand{\vxi}{\boldsymbol{\xi}}
\newcommand{\Dvxi}{{\cal{D}}_{ x,\alpha}( \vxi)}
\newcommand{\Dvxip}{{\cal{D}}_{ x',\gamma}( \vxi')}
\newcommand{\Dvxii}{{\cal{D}}_{ x_i,\alpha_i}( \vxi_{i})}
\newcommand{\Dvxig}{{\cal{D}}_{ x,\gamma}( \vxi)}
\newcommand{\Dvxigp}{{\cal{D}}_{ x',\gamma}( \vxi')}

\newcommand{\Dxix}{{\cal{D}}_{ x,\alpha}( \xi_x)}
\newcommand{\roa}{\rho_{\alpha}}
\newcommand{\roab}{\rho_{\alpha}^{\scriptscriptstyle B}}
\newcommand{\Lr}{{\mathcal{L}}}
\newcommand{\calV}{{\mathcal{V}}}

\newcommand{\vcm}{{\cal {V}}^{{\rm c \, m }}}
\newcommand{\vmc}{{\cal {V}}^{{\rm m \, c }}}
\newcommand{\vmm}{{\cal {V}}^{{\rm m \, m }}}
\newcommand{\Fcc}{F^{\rm c \, c}}
\newcommand{\fcc}{f^{\rm c \, c}}
\newcommand{\Fcm}{F^{\rm c \, m}}
\newcommand{\fcm}{f^{\rm c \, m}}
\newcommand{\Fmc}{F^{\rm m \, c}}
\newcommand{\fmc}{f^{\rm m \, c}}
\newcommand{\Fmm}{F^{\rm m \, m}}
\newcommand{\fmm}{f^{\rm m \, m}}
\newcommand{\ftt}{f_{\scriptscriptstyle{\rm TT}}}
\newcommand{\Frt}{F_{\scriptscriptstyle{\rm RT}}}
\newcommand{\cg}{{\mathbb{G}}}

\newcommand{\cpps}{{\mathbb{P}}^{\star}}
\newcommand{\zs}{z^{\rm sc}}
\newcommand{\calVs}{{\cal {V}}^{\rm sc}_{\rm cloud}}
\newcommand{\Vs}{V^{\rm sc}_{\rm cloud}}
\newcommand{\kbar}{{\overline{\kappa}}}
\newcommand{\kz}{\kappa}
\newcommand{\kd}{\kappa_{{\scriptscriptstyle D}}}
\newcommand{\vy}{{\bf y}}
\newcommand{\vyt}{\widetilde{{\bf y}}}
\newcommand{\vq}{{\bf q}}
\newcommand{\vk}{{\bf k}}
\newcommand{\phit}{{\widetilde{\phi}}}

\newcommand{\xt}{{\widetilde{x}}}
\newcommand{\klat}{{{\widetilde{\lambda}}_{\alpha}}}
\newcommand{\vecrt}{{\widetilde{\vecr}}}
\newcommand{\rac}{\sqrt{1+\vq^2}}
\newcommand{\ract}{\sqrt{t^2-1}}
\newcommand{\phitzero}{{\widetilde{\phi}}^{(0)}}
\newcommand{\cO}{{\cal O}}
\newcommand{\phizero}{\phi^{(0)}}
\newcommand{\eps}{\varepsilon}
\newcommand{\epsd}{\varepsilon_{\scriptscriptstyle D}}
\newcommand{\epsa}{\varepsilon_{\alpha}}
\newcommand{\kzl}{\kz \lambda}
\newcommand{\kdl}{\kd \lambda}
\newcommand{\Lb}{{\overline{L}}}
\newcommand{\Vszero}{V^{{\rm sc} (0)}_{\rm cloud}}
\newcommand{\eg}{e_{\gamma}}
\newcommand{\zg}{z_{\gamma}}

\newcommand{\epsg}{\varepsilon_{\gamma}}

\newcommand{\zsc}{z^{{\rm sc}}}
\newcommand{\zbsca}{{\overline{z}}^{\rm sc}_{\alpha}}
\newcommand{\zbscg}{{\overline{z}}^{\rm sc}_{\gamma}}
\newcommand{\diagfcc}{\mbox{\begin{picture}(45,25)(13,-3)
\put(20,0){\circle{10}}
\put(25,0){\line(1,0){20}}
\put(50,0){\circle{10}}
\put(30,10){$\fcc$}
\end{picture}}}
\newcommand{\diagf}{\mbox{\begin{picture}(45,25)(13,-3)
\put(20,0){\circle{10}}
\put(25,0){\line(1,0){20}}
\put(50,0){\circle*{10}}
\put(35,10){$f$}
\end{picture}}}
\newcommand{\diagfmc}{\mbox{\begin{picture}(45,25)(13,-3)
\put(20,0){\circle{10}}
\put(45,0){\vector(-1,0){20}}
\put(50,0){\circle{10}}
\put(30,10){$\fmc$}
\end{picture}}}
\newcommand{\diagFcc}{\mbox{\begin{picture}(45,25)(13,-3)
\put(20,0){\circle{10}}
\put(25,0){\line(1,0){20}}
\put(50,0){\circle*{10}}
\put(30,10){$\Fcc$}
\end{picture}}}
\newcommand{\diagbFcc}{\mbox{\begin{picture}(45,25)(13,-3)
\put(20,0){\circle{10}}
\put(25,0){\line(1,0){20}}
\put(50,0){\circle{10}}
\put(30,10){$\Fcc$}
\end{picture}}}

\newcommand{\diagbFccd}{\mbox{\begin{picture}(45,25)(13,-3)
\put(20,0){\circle{10}}
\put(24.5,-2){\line(1,0){21}}
\put(24.5,2){\line(1,0){21}}
\put(50,0){\circle{10}}
\put(22,10){$\frac{1}{2} [\Fcc]^2$}
\end{picture}}}
\newcommand{\diagbfccd}{\mbox{\begin{picture}(45,25)(13,-3)
\put(20,0){\circle{10}}
\put(24.5,-2){\line(1,0){21}}
\put(24.5,2){\line(1,0){21}}
\put(50,0){\circle{10}}
\put(22,10){$\frac{1}{2} [\fcc]^2$}
\end{picture}}}

\newcommand{\diagbFmc}{\mbox{\begin{picture}(45,25)(13,-3)
\put(20,0){\circle{10}}
\put(45,0){\vector(-1,0){20}}
\put(50,0){\circle{10}}
\put(30,10){$\Fmc$}
\end{picture}}}

\newcommand{\diagdfcc}{\mbox{\begin{picture}(65,25)(13,-3)
\put(20,0){\circle{10}}
\put(25,0){\line(1,0){40}}
\put(45,0){\circle*{10}}
\put(70,0){\circle{10}}
\put(54,10){$\fcc$}
\put(30,10){$\fcc$}
\put(44,-15){$z$}
\end{picture}}}
\newcommand{\diagdf}{\mbox{\begin{picture}(65,25)(13,-3)
\put(20,0){\circle{10}}
\put(25,0){\line(1,0){40}}
\put(45,0){\circle*{10}}
\put(70,0){\circle*{10}}
\put(57,10){$f$}
\put(32,10){$f$}
\put(43,-15){$z$}
\put(68,-15){$z$}
\end{picture}}}
\newcommand{\diagtf}{\mbox{\begin{picture}(55,25)(13,-3)
\put(20,0){\circle{10}}
\put(25,0){\line(2,1){25}}
\put(25,0){\line(2,-1){25}}
\put(50,-10){\line(0,1){25}}
\put(50,12){\circle*{10}}
\put(50,-13){\circle*{10}}
\put(58,-2){$f$}
\put(32,11){$f$}
\put(32,-16){$f$}
\put(48,-26){$z$}
\put(48,21){$z$}
\end{picture}}}
\newcommand{\diagfccdfcc}{\mbox{\begin{picture}(55,25)(13,-3)
\put(20,0){\circle{10}}
\put(60,0){\circle{10}}
\put(25,0){\line(1,0){30}}
\put(25,0){\line(2,1){15}}
\put(55,0){\line(-2,1){15}}
\put(40,8){\circle*{10}}
\put(21,11){$\fcc$}
\put(49,11){$\fcc$}
\put(35,-13){$\fcc$}
\put(37,19){$z$}
\end{picture}}}
\newcommand{\diagdfccdfcc}{\mbox{\begin{picture}(55,25)(13,-3)
\put(20,0){\circle{10}}
\put(60,0){\circle{10}}
\put(25,0){\line(2,1){15}}
\put(55,0){\line(-2,1){15}}
\put(25,0){\line(2,-1){15}}
\put(55,0){\line(-2,-1){15}}
\put(40,8){\circle*{10}}
\put(40,-7){\circle*{10}}
\put(21,11){$\fcc$}
\put(49,11){$\fcc$}
\put(21,-17){$\fcc$}
\put(49,-17){$\fcc$}
\put(37,19){$z$}
\put(37,-20){$z$}
\end{picture}}}
\newcommand{\diagfcctfcc}{\mbox{\begin{picture}(45,25)(13,-3)
\put(20,0){\circle{10}}
\put(20,30){\circle*{10}}
\put(50,30){\circle*{10}}
\put(50,0){\circle{10}}
\put(25,0){\line(1,0){20}}
\put(25,30){\line(1,0){20}}
\put(50,5){\line(0,1){20}}
\put(20,5){\line(0,1){20}}
\put(4,13){$\fcc$}
\put(55,13){$\fcc$}
\put(30,-12){$\fcc$}
\put(30,35){$\fcc$}
\put(20,40){$z$}
\put(50,40){$z$}
\end{picture}}}
\newcommand{\diagfmcfcc}{\mbox{\begin{picture}(65,25)(13,-3)
\put(20,0){\circle{10}}
\put(45,0){\vector(-1,0){20}}
\put(25,0){\line(1,0){40}}
\put(47,0){\circle*{10}}
\put(70,0){\circle{10}}
\put(54,10){$\fcc$}
\put(30,10){$\fmc$}
\put(44,-15){$z$}
\end{picture}}}
\newcommand{\diagtfcc}{\mbox{\begin{picture}(85,25)(13,-3)
\put(20,0){\circle{10}}
\put(25,0){\line(1,0){60}}
\put(43,0){\circle*{10}}
\put(67,0){\circle*{10}}
\put(90,0){\circle{10}}
\put(30,10){$\fcc$}
\put(54,10){$\fcc$}
\put(77,10){$\fcc$}
\put(42,-15){$z$}
\put(66,-15){$z$}
\end{picture}}}
\newcommand{\diagfmcdfcc}{\mbox{\begin{picture}(85,25)(13,-3)
\put(20,0){\circle{10}}
\put(45,0){\vector(-1,0){20}}
\put(25,0){\line(1,0){60}}
\put(45,0){\circle*{10}}
\put(68,0){\circle*{10}}
\put(90,0){\circle{10}}
\put(30,10){$\fmc$}
\put(54,10){$\fcc$}
\put(77,10){$\fcc$}
\put(42,-15){$z$}
\put(65,-15){$z$}
\end{picture}}}
\newcommand{\B}{{\cal{B}}}

\newcommand{\Mb}{{\overline{M}}}
\newcommand{\De}{\Delta_{{\rm el}}}
\newcommand{\ew}{\epsilon_{{\scriptscriptstyle W}}}

\newcommand{\dxdslad}{2 x^2 \,/ \, \laa^2}
\newcommand{\Pas}{\Phi_{{\rm as}}}

\newcommand{\Phiclu}{\Phi^{{\rm cl} (\eps)}}
\newcommand{\Phiquu}{\Phi^{{\rm qu} (\kzl)}}
\newcommand{\somg}{\sum_\gamma}
\newcommand{\rogb}{\rho_\gamma^{{\scriptscriptstyle B}}}
\newcommand{\Pucl}{A}
\newcommand{\Erfc}{{\rm{Erfc}}}
\newcommand{\Erf}{{\rm{Erf}}}

\newcommand{\Ir}{I_{{\rm r}}}

\newcommand{\kb}{k_{{\scriptscriptstyle B}}}
\newcommand{\plsap}{\left( \lambda / a \right)}

\begin{document}

\thispagestyle{empty}

\title{\bf Density profiles in a quantum Coulomb fluid near a hard wall}

\author{{\bf J.-N. Aqua}\thanks{Laboratoire de Physique
(Laboratoire associ\'e au Centre National de la Recherche Scientifique -- UMR 5672), \'Ecole Normale Sup\'erieure de Lyon, 46 allée d'Italie, 69364 Lyon 
\textsc{France}.
Present address : Institute for Physical Science and Technology, University of Maryland, College Park, Maryland 20910, \textsc{USA}.} \hspace{0.1cm} {\bf  and F. Cornu}\thanks{Laboratoire de Physique Th\'eorique
(Laboratoire associ\'e au Centre National de la Recherche Scientifique - UMR 8627), 
Universit\'e Paris-Sud, B\^{a}timent 210, 91405 Orsay, \textsc{France}.}}

\date{\today}

\maketitle 

\begin{abstract}

  Equilibrium particle densities near  a hard wall are studied
for a quantum fluid made of point charges which interact 
via Coulomb potential without any regularization. 
In the framework of the grand-canonical 
ensemble, we use an equivalence   with a classical system
of loops with random shapes, based on the  Feynman-Kac path-integral representation of the quantum Gibbs factor. After systematic resummations of Coulomb divergences in the
 Mayer fugacity expansions of  loop densities, 
there appears a  screened potential $\phi$. It obeys an inhomogeneous Debye-H\"uckel 
equation with an effective screening length which depends on the distance from the wall. The formal solution for $\phi$ can be expanded in powers of the ratios of the de Broglie thermal wavelengths $\laa$'s of each species $\alpha$ and the limit of the screening length  far away from the wall. 
In a regime of low degeneracy and weak coupling,
exact analytical density profiles are calculated at first order in two independent parameters. Because of the vanishing of wave-functions close to the wall, density profiles vanish gaussianly fast in the vicinity of the wall over  distances 
$\laa$'s, with an essential singularity in Planck constant $\hbar$. 
When  species have different masses,  this effect is equivalent to the appearance of a  quantum surface charge  localized on the wall and proportional to  
$\hbar$ at leading order.  Then,  density profiles, as well as the electrostatic potential drop created by the charge-density profile, also involve a term linear in $\hbar$  and
which decays exponentially fast  over the classical Debye 
screening length $\xid$. The corresponding contribution to the global surface charge exactly compensates the  charge in the very vicinity of  the surface, so that 
  the net electric field vanishes in the bulk, as it should.

{\bf KEYWORDS~:} Coulomb interactions, quantum mechanics, \linebreak hard wall, grand-canonical ensemble, inhomogeneous Debye equation, surface charge. 
\vskip 1cm 

\end{abstract}

\numberwithin{equation}{section}

\section{Introduction}

\subsection{Issue at stake}

In the present  paper the equilibrium density profiles in a quantum  fluid of point charges are studied in the vicinity of an impenetrable hard wall. The wall, which occupies the semi-infinite  region $x<0$, has no internal structure and its dielectric constant is the same as that of the medium where charges move. On the contrary, the fluid made of $n_s$ particle species is described at the microscopic level in the framework of quantum statistical mechanics. Two  point charges $\ea$ and $\eap$  (where $\alpha$ is a species index) interact via the electrostatic 
 interaction   
$\ea \eap \, v(\vecr - \vecr')$, where 
\begin{equation}
  \label{defv}
  v(\vecr - \vecr') = \frac{1}{|\vecr - \vecr'|}
\end{equation}
in Gauss units.
(The charge $\ea$ includes a factor $1/\sqrt{\epsilon_{\rm m}}$ in energy terms when
charges are embedded in a continuous medium with a relative  dielectric constant $\epsilon_{\rm m}$ with respect to the vacuum.)
The interaction is translationally invariant, and the anisotropy lies only in the geometric constraint enforced by the presence of the wall.
The exact analytical expressions of the density profiles 
$\roa(x)$'s are obtained in a regime where
 exchange effects are negligible and where Coulomb coupling is weak. Results hold for the electron-hole gas in an intrinsic semi-conductor in the vicinity of a junction or for a dilute and hot quantum plasma near a vessel wall.

The interesting point  of the model is that it exhibits 
how a quantum charge effect, gaussianly localized  over 
de Broglie thermal wavelengths  
in the vicinity of the wall,  is carried by long-range  Coulomb  interactions up to
larger distances from the wall, with an exponential decay over a scale equal to the coulombic screening length. 
Indeed, whereas the density in a classical ideal gas is uniform in the whole region $x>0$ and is discontinuous on the wall surface, the quantum density is continuous and vanishes at 
$x=0$, because of the continuity  of wave-functions and  their cancellation inside the impenetrable wall. At the inverse temperature $\beta = 1 / \kb T$ (where $\kb$ is Boltzmann constant),  in a low-degeneracy limit quantum statistics is reduced to Maxwell-Boltzmann statistics, and the density $\roa^{\rm id}(x) $ of species $\alpha$ in an ideal gas with quantum dynamics vanishes gaussianly fast over the scale of the 
thermal de Broglie wavelength $\laa $,
\begin{equation} 
\label{rhoid}
\roa^{\rm id}(x) =\roab
\left(1 - e^{- 2 x^2/\laa^2}\right)
+\roab\cO\left(\left(\frac{\laa}{a_{\alpha}}\right)^3\right).
\end{equation}
In \eqref{rhoid}
  $\roab$
is the  bulk  density for species $\alpha$, and 
\begin{equation}
\laa = \hbar \sqrt{\frac{\beta }{\ma}},
\end{equation} 
where $\hbar$ is Planck constant and $\ma$ is the mass of species $\alpha$.
The quantum expression with Maxwell-Boltzmann statistics \eqref{rhoid} is valid up to terms of order 
$(\laa/a_{\alpha})^3$,
 where $a_{\alpha}$ is 
the  mean interparticle distance between particles of the same species $\alpha$ ($(4/3)\pi \roab a_{\alpha}^3=1$). 

When Coulomb interactions are taken into account, there arises only one third typical length scale, because 
 Coulomb interactions are scale-invariant. Then
only
 two independent dimensionless parameters rule the  physical regimes of the system: the degeneracy parameter and the Coulomb coupling parameter.
The degeneracy parameter  is $(\lambda/a)^3$, where $\lambda = \sup_\alpha \{ \laa \}$ and $a$ is a typical mean interparticle distance.
 When exchange effects are negligible,
\begin{equation}
  \label{lsa}
 \left(\frac{\lambda}{a}\right)^3 \ll 1.
\end{equation}
 The third length scale may be chosen to be either the classical screening length $\xid$
\begin{equation}
\label{defxid}
\xid^{-1}=\kd\equiv \sqrt{4\pi \beta \sum_{\alpha=1}^{n_s}
\roab \ea^2}
\end{equation} 
 or the Landau length 
$\beta e^2$, namely the classical closest approach distance  
 between two typical like-charges 
$e$ with kinetic energy of order $1/\beta$.
Henceforth, the classical Coulomb coupling parameter can be chosen to be equal either to the ratios $(a/\xid)^3$,
 $ \beta e^2/a\equiv\Gamma$, or   
$  \beta e^2/\xid\equiv 2\epsd$.
 These ratios are proportional to one another,
\begin{equation}
\label{rels}
\left(\frac{a}{\xid}\right)^3 \propto \epsd
\propto \Gamma^{3/2}.
\end{equation}
More precisely they are linked by the relations $(a/\xid)^3= C\epsd$ and $\epsd=[C/8]^{1/2} \,\Gamma^{3/2}$, where $C$ is a numerical factor 
which depends on the composition of the fluid through $\xid$ and on the relation between $a$ and  bulk densities. 
In a   weak-coupling regime
\begin{equation}
 \label{petitG} 
\left(\frac{a}{\xid}\right)^3  \ll 1.
\end{equation}
When both \eqref{lsa} and \eqref{petitG} are satisfied, 
\begin{equation}
\label{encadra}
\lambda\ll a \ll\xid.
\end{equation} 
Therefore, the  low-degeneracy  and weak-coupling  regime is also a regime where 
$\kd \lambda \ll 1$.

\subsection{Results}

In the low-degeneracy  and weak-coupling  regime defined by \eqref{lsa} and \eqref{petitG}, the analytical expression for the profile density $\roa(x)$ is calculated in a subregime where the first coupling correction, of order $\epsd=\kd \beta e^2/2$, and the first diffraction correction, of order $\kd \lambda$, dominate other coupling and exchange corrections. 
 At   order  $\epsd$ classical contributions do not involve the short-range cut-off that must be introduced in order to prevent the collapse of the system in the limit where $\hbar$ tends to zero. As shown in Section \ref{weak}, the subregime corresponds to a scaling where
\begin{equation}
\label{regimed}
\epsd^2\leq \left( \frac{\lambda}{a}\right)^3\ll \epsd.
\end{equation}
\eqref{regimed} can be reexpressed as
$\epsd^3\leq (\kdl)^3\ll \epsd^2$,
since $(\lambda/a)\propto\kdl /\epsd^{1/3}$ by virtue of \eqref{rels}.
In the subregime \eqref{regimed} the density profile in the region $x>0$ reads
\begin{eqnarray}
  \label{qdensfnch}
&&  \roa (x) =\roab\left( 1 - e^{- \dxdslad} \right)
\nonumber\\ 
&&\qquad
\times  \left[1- \frac{1}{2} \kd \beta \ead \, \Lb (\kd x) 
-\beta\ea \Phi(x)\right] +
\roab \,\, o\left(\epsd,\kdl\right)
\end{eqnarray}
where $o\left(\epsd,\kdl\right)$ denotes  a sum of  terms which tend to zero faster than either $\epsd$ or $\kdl$  when these parameters vanish (See \eqref{defetad}). When the latter terms are neglected, 
$\roa (x)$ appears as the product of the ideal-gas density  
\eqref{rhoid} with a function arising from interaction corrections. (The generic properties of density profiles are discussed in Section \ref{prop}.)

The density profile \eqref{qdensfnch} results from the combination of three effects: first,  the vanishing of quantum wave-functions in the vicinity of the wall; second, the geometric repulsion from the wall \cite{jn&fr01I}, described by 
the classical part of the screened self-energy due to the deformation of screening clouds, $(1/2)\kd \beta \ead \, \Lb (u)$,  given in \eqref{defLbq}; 
 third, the interaction  $\ea \Phi(x)$ with the electrostatic potential drop  $\Phi(x)$ with respect to the bulk, which is  created by the charge density profile $\sum_{\gamma}
\eg \rho_{\gamma}(x)$ itself. (The sign of the latter interaction  depends on the sign of $\ea$.) 
The potential drop
$\Phi(x)$ in \eqref{qdensfnch}
is the sum of a classical contribution \cite{jn&fr01I} and a quantum ``{diffraction}'' effect, linear in $\hbar$,
\begin{equation}
  \label{}
  \Phi (x) = \Phi^{{\rm cl}(\epsd)}(x)+
\Phi^{{\rm qu}(\kdl)} (x). 
\end{equation}
$\Phi^{{\rm cl}(\epsd)}(x)$, of order $\epsd/(\beta e)$, is written in \eqref{rappelPhicl} and  
$\Phi^{{\rm qu}(\kdl)} (x)$, of order $\kdl/(\beta e)$, reads
\begin{equation}
  \label{valuePhiqu}
\Phi^{{\rm qu}(\kdl)} (x) = - \hbar \B e^{-\kd x}\quad 
\mbox{with $\displaystyle\B= \frac{\pi}{\sqrt{2}}
\frac{\somg (\eg / \sqrt{\mg})\rogb}{\sqrt{\soma 
\ead \roab}}$}.
\end{equation}
$\Phi^{{\rm qu}(\kdl)} (x)$ appears only when species  have different masses, because of the  bulk local neutrality 
\begin{equation}
  \label{neutrbulk}
  \sum_{\alpha} \ea \roab = 0.
\end{equation}

The density profile 
\eqref{qdensfnch} can be rewritten as
\begin{equation}
\label{factcl}
\roa(x)=
\left(1-e^{-2x^2/\laa^2}\right)
\left[\roa^{{\rm cl}(\epsd)}(x)+\hbar \roab \beta \ea  \B e^{-\kd x}\right]
+\roab \, \,o\left(\epsd,\kdl\right),
\end{equation} 
where $\roa^{{\rm cl}(\epsd)}(x)$ is the classical density profile 
calculated up to relative order $\epsd$ \cite{jn&fr01I}  and  
written in \eqref{valuerhocl}. 
We stress that the direct contribution \eqref{rhoid} from the vanishing of wave-functions in the ranges  $\laa$'s from the wall has an essential singularity in $\hbar$, whereas 
the quantum 
part   of the electrostatic potential  
at leading order, $\Phi^{{\rm qu}(\kdl)} (x) $,  is linear in $\hbar$. 
(We recall that for systems invariant under translations -- which is not the case here -- and with sufficiently smooth potentials -- such as the Coulomb interaction --  $\hbar$-expansions involve only even powers of $\hbar$, as can be seen for instance in
Wigner-Kirkwood expansions.)

The appearance in the electrostatic potential
$\Phi(x)$ 
of a $\hbar$-term which decays exponentially fast over the classical Debye screening length $\xid$ has the following physical interpretation. 
When  species have different masses, the global
 charge $\sigma_{<}$  carried by the fluid (per unit area) over all  distances  $x<a$ from the wall is essentially created at leading order by the differences in the Gaussian density profiles 
and  is concentrated  over a width  of  order $\lambda$. 
Since $\lambda$ is negligible with respect to the bulk mean interparticle distance $a$, the leading-order charge $\sigma^{{\rm qu}(\kdl)}_{<}$ can be seen as a surface charge localized at $x=0$.
 As shown in Section \ref{section64}, the surface charge $\sigma^{{\rm qu}(\kdl)}_{<}$, which appears even in the zero-coupling limit,  creates an electrostatic potential  through the classically-screened Coulomb  interaction (calculated at leading order), and  this potential is equal to the leading $\hbar$-term 
$\Phi^{{\rm qu}(\kdl)}(x)$ in the electrostatic potential $\Phi(x)$ created by the charge-density profile $\soma \ea\roa(x)$. Moreover $\Phi^{{\rm qu}(\kdl)}(x)$ is involved in the density profiles in such a way that the leading-order global charge $\sigma^{{\rm qu}(\kdl)}_{>}$  carried by the fluid (per unit area) over all distances $ x>a$, and which is dilute over the scale $\xid$, compensates 
$\sigma^{{\rm qu}(\kdl)}_{<}$ (see Section \ref{section64}).
Indeed, since the wall is made of an insulating material and  carries no external charge,
the global surface charge    $\sigma$ carried by the
fluid per unit area vanishes at equilibrium \cite{Janco86} 
\begin{equation}
\label{neutrsurf}
\sigma \equiv \int_0^{\infty} dx \, \soma \ea \roa (x) = 0.
\end{equation}
 $\sigma^{{\rm qu}(\kdl)}_{<}$ and $\Phi^{{\rm qu}(\kdl)}(x=0)$ are estimated in the case of the intrinsic semiconductor GaSb.
The case where the wall has not the same dielectric constant as the medium where the fluid is embedded is commented in Section \ref{section7}.

\subsection{Methods}

Before going into details, we summarize the general methods displayed in Sections \ref{form}--\ref{weak}.

First, 
a formalism based on path integrals and 
devised for the study of bulk properties in Coulomb fluids -- with Maxwell-Boltzmann statistics \cite{ACP94} then quantum statistics 
\cite{Cornu96I}-- is generalized to a semi-infinite geometry (Section \ref{form}). The system is studied in the grand-canonical ensemble (Section \ref{section21}).
A degeneracy of physical quantities with respect to fugacities arises from the neutrality constraints enforced by the long-range of Coulomb interactions.
We investigate the nature of this degeneracy, and we show that we are allowed to split the  latter degeneracy in order to impose the local neutrality in the zero-coupling limit (Section \ref{section22}). (This trick allows to simplify  weak-coupling expansions performed in Section \ref{weak}.) By use of the Feynman-Kac formula (Section \ref{section23}), quantum dynamics can be described by a functional integral over Brownian paths, which correspond to quantum position fluctuations. As in the bulk situation, 
the quantum system of point charges is  equivalent to a classical
system of loops with random shapes (Section \ref{section25}). The only difference in formulae for the bulk or for  the vicinity of the wall is that the path measure is  anisotropic and depends on the distance from the wall in the second case. 

Then methods originally devised for classical fluids with  internal degrees of freedom can be used (Section \ref{diag}). In Section \ref{section31} we introduce generalized Mayer diagrams for the fugacity expansion of the loop density of each species. Point weights in those diagrams depend both on the internal degrees of freedom of loops -- charge and  shape -- and on the distance $x$ from the wall.
Because of the long range of Coulomb interaction, every Mayer diagram that is not sufficiently connected corresponds to a divergent integral in the thermodynamical limit. These divergences disappear after exact systematic resummations analogous to those performed  in Ref.\cite{Cornu96I} (Section \ref{section32}). (Details are provided in Appendix \ref{appresum}.)  Resummations introduce  
  a screened potential 
$\phi (\vecr, \vecr')$, solution of an inhomogeneous Debye equation 
\begin{equation}
  \label{eqgene}
 \left[\Delta_{\vecr} - \kbar^2(x)\right] \phi(\vecr,\vecr') =-4\pi \delta
 (\vecr-\vecr'),
\end{equation}
where the effective screening length $1/\kbar(x)$ depends on the 
distance $x$ to the wall because of the  vanishing of 
wave-functions at the wall surface (see
\eqref{rhoid}).

At this point the difficulty to be circumvented is  the resolution of  equation \eqref{eqgene}
 (Section \ref{ldlim}).
The equation can be turned into a one-dimensional differential equation by considering the Fourier transform \linebreak
$\phi(x,x',\vk)$ of $\phi(x,x',\vy)$ in the directions parallel to the wall surface (Section \ref{section33}). Let 
$\phi^{(0)}(x,x',\vy)$ be the expression that $\phi(x,x',\vy)$ would take if the profile $\kbar(x)$ were uniform 
and equal to its bulk value $\kz$ in the region $x>0$.
A formal series representation of the solution $\phi(x,x',\vk)-\phi^{(0)}(x,x',\vk)$
 has been given  in Ref. 
\cite{jn&fr01II}, where a similar equation arises in the case of a classical charge fluid in the vicinity of a wall with an electrostatic response.  
This series provides an expansion of the solution  $\phi(x,x',\vk)$ around $\phi^{(0)}(x,x',\vk)$ in powers of the small parameter $\kz \lambda$, where $\kz$ is the limit of
$\kbar (x)$ when $x$ goes to infinity, while $\lambda$ is the length scale over which $\kbar (x)$ varies quickly when $x$ approaches $0$. The expansion  of  $\phi(x,x',\vk)-\phi^{(0)}(x,x',\vk)$ in powers of $\kz \lambda$ is uniform in $x$ and $x'$.

 In the low-degeneracy  and weak-coupling  regime to be studied
(Section \ref{section51}), the condition $\kz \lambda\ll 1$ is met. The screened self-energy is purely classical at leading order and the quantum correction appears only at order $\eps\times \kzl$, where $\eps$ is defined as $\epsd$ with $\kz$ in place of $\kd$, $\eps\equiv(1/2)\kz \beta e^2$
(Section \ref{section43}).
 A scaling
analysis performed in the low-degeneracy and weak-coupling limit
(Section \ref{section52} and Appendix \ref{appscal})
shows that only one resummed Mayer   diagram  contributes to density profiles at first 
order in  $\eps$ and $\kzl$. The electrostatic potential drop $\Phi(x)$ created by the charge-density profile is identified in the formal expression of the contribution from this diagram (Section \ref{section53}). Because of the local neutrality condition in the bulk, only 
 the   classical zeroth-order term in the $\kz \lambda$-expansion of $\phi-\phi^{(0)}$ proves to contribute  to the potential drop $\Phi(x)$ at leading orders $\eps$ and $\kzl$.


\section{General formalism}
\label{form}

\subsection{Grand-canonical ensemble and statistics}
\label{section21}

We recall that we consider a fluid made of $n_s$ species (indexed by $\alpha$), each of
which is characterized by its mass $\ma$, its charge $\ea$, and  its spin $S_{\alpha}\hbar$. (In the following, interactions involving spins will be neglected and spin will only determine the nature of quantum statistics.) 
 The Hamiltonian operator
$\widehat{H}_{\{N_\alpha\}}$ of a system which contains 
$N_{\alpha}$ particles of each species $\alpha$
reads
\begin{equation}
  \label{Hquant}
  \widehat{H}_{\{N_\alpha\}} = \sum_i \frac{{\widehat {\bf p}}_{i}^{\, 2}}
  {2 m_{\ai}} +  \sum_i \widehat{\Vsr}(x_i)+
  \sum_{i < j} e_{\ai} e_{\aj} \widehat{\vb} 
  (\vecr_i -\vecr_j).
\end{equation}
($\{N_{\alpha}\}$ is a shorthand notation  for 
$\{N_{\alpha}\}_{\alpha=1,\ldots,n_s}$ and the particle index $i$ runs from $1$
to $N=\soma N_\alpha$.) In \eqref{Hquant} the first term which
involves the momentum operator ${\widehat {\bf p}}$ is
the global kinetic energy of the system. The second term is a sum of one-body
potentials $\widehat{\Vsr}(x_i)$ which describe the particle-wall interactions. We choose a
simple classical hard-wall modelization, where the atomic structure of the wall is ignored. The effect of the wall is only  to prevent
particle wave-functions from propagating inside the 
negative-$x$ region 
occupied by the wall, 
\begin{equation}
\label{defVsr}
 \Vsr(x) = \left\{ \begin{array}{ll}
                    +\infty  & \mbox{if $x<0$}\\                      
                    0 & \mbox{if $x>0$.}
                     \end{array}
                   \right.        
\end{equation}
The wall repulsion is independent of the particle species. The sum of
pair interactions in the third term 
involves only Coulomb potential \eqref{defv}.


The fixed parameters of the system are the volume $\vert\Lambda\vert$ of the region $\Lambda$ that the fluid occupies, the area $\vert\partial_{\scriptscriptstyle W}\Lambda\vert$ of the fluid-wall interface,
the temperature, and the densities $\roab$'s far away from the boundaries of 
 $\Lambda$. We use the grand-canonical ensemble where the  parameters are
 the volume $\vert\Lambda\vert$, the area $\vert\partial_{\scriptscriptstyle W}\Lambda\vert$, the inverse temperature $\beta$, and the chemical 
potentials $\{\mua(V_{\scriptscriptstyle R})\}_{\alpha=1,\ldots,n_s}$ of particles in a reservoir where the electrostatic potential takes the uniform value $V_{\scriptscriptstyle R}$. The 
 grand partition function reads
 \begin{equation}
  \label{defuXi}
  \Xi(\beta, \{\mua\}, \vert\Lambda\vert, \vert\partial_{\scriptscriptstyle W}\Lambda\vert) 
  = \sum_{\{N_\alpha\}} 
  {\rm Tr}_{\Lambda, \{N_\alpha\}}^{\rm sym}  \, e^{- \beta 
\left[\widehat{H}_{\{N_\alpha\}}-\soma\mua\widehat{N}_{\alpha}
\right]},
\end{equation}
where the trace ${\rm Tr}_{\Lambda, \{N_\alpha\}}^{\rm sym}$ 
is restricted to the quantum 
states that are properly symmetrized according to the Bose-Einstein  or Fermi-Dirac statistics obeyed by each
species. ($\widehat{N_{\alpha}}$ is the particle-number operator for species
$\alpha$.)
 As in the classical case,
the density profile $\roa (x)$ can be determined by using a functional derivation  of  
$\Xi \left[ \widetilde{\mua} \right]$, where 
$\sum_{\alpha}\mua{\widehat N}_\alpha$ is replaced by 
$\int_{\Lambda} d\vecr \widetilde{\mua}(x) {\widehat \roa}(x)$.
The relation is 
\begin{equation}
\label{relrhoXi}
\roa (x) =\lim_{\vert\Lambda\vert\rightarrow +\infty} 
\frac{1}{\beta }  \, \left. \frac{\delta
 \ln \Xi \left[ \widetilde{\mua} \right]}
{\delta \widetilde{\mua} (x)}  
\right|_{\widetilde{\mua} (x) = \mua }.
\end{equation}


The formalism and  results presented in Sections \ref{form}  and \ref{diag}  can be obtained
with quantum statistics (as detailed in Section \ref{section25}). However, our aim is to produce explicit analytical
results in the low-degeneracy regime \eqref{lsa}. We have checked 
that quantum statistics
effects arise only at order $(\lambda/a)^3$. In other words \cite{ACP94}
\begin{equation} 
\label{relXiXiMB}
  \Xi =  \Xi_{\rm MB}
  +\cO\left(\left(\frac{\lambda}{a}\right)^3\right),
\end{equation}
 where the Maxwell-Boltzmann grand partition function $\Xi_{\rm MB}$ is a trace over tensorial products of one-particle wavefunctions which are not symmetrized according to species statistics.
 If 
 the tensorial product of $N=\soma N_{\scriptscriptstyle \alpha}$
one-particle states in position representation is denoted by $\vert \{ \vecr_i \} \rangle $, the grand partition function $\Xi_{\rm MB}$,
where only dynamics is quantum, reads
\begin{eqnarray}
  \label{defXi}
 && \Xi_{\rm MB}(\beta, \{\mua\}, \vert\Lambda\vert,
\vert\partial_{\scriptscriptstyle W}\Lambda\vert) \\
 &&\quad = \sum_{\{N_\alpha\}} \left[\prod_{\alpha}  \frac{e^{\beta\mua N_\alpha}(2S_{\alpha}+1)^{N_\alpha}}{N_\alpha!} \right]
 \int   \left[\prod_{i=1}^{N}  d\vecr_{i}\right]
\langle \{ \vecr_{i} \}
 \vert e^{- \beta \widehat{H }_{\{N_\alpha\}} }\vert \{ \vecr_i \}
  \rangle.\nonumber   
\end{eqnarray}
where $2S_{\alpha}+1$ is the spin degeneracy factor, which arises  because spin   does not appear in the expression \eqref{Hquant} of the
Hamiltonian.  
In \eqref{defXi} we have used the commutativity of the operators
$\widehat{H}_{\{N_\alpha\}}$ and ${\widehat{N}}_\alpha$'s.


\subsection{Degeneracy with respect to fugacities}
\label{section22}

In the following we will take advantage of a degeneracy of physical quantities with respect to fugacities that arises from the vanishing of the
global volumic and surfacic charges of the system  in the thermodynamic limit. 
If $\langle\cdots\rangle$ denotes a grand-canonical average, the thermodynamic limit of the  charge in the fluid  is
\begin{equation}
\label{chargemacro}
\lim_{\rm th}\langle\sum_{\alpha} \ea N_{\alpha}\rangle
=
\left(\sum_{\alpha} \ea \roab\right)\vert\Lambda\vert
+\sigma \vert\partial_{\scriptscriptstyle W}\Lambda\vert
+o\left(\vert\partial_{\scriptscriptstyle W}\Lambda\vert\right),
\end{equation} 
where $o\left(\vert\partial_{\scriptscriptstyle W}\Lambda\vert\right)$ denotes a term which diverges more slowly than the area 
$\vert\partial_{\scriptscriptstyle W}\Lambda\vert$ when the size of the domain $\Lambda$ becomes infinite. The expression of $\sigma$ in terms of the thermodynamic limits of density profiles is given in
\eqref{neutrsurf}.
As a consequence of  the existence of the thermodynamical limit 
\cite{Lieb&Lebo72}, the macroscopic volumic charge 
$\left(\sum_{\alpha} \ea \roab\right)\vert\Lambda\vert$ vanishes, and, in the case of an insulating  hard wall  that 
is not externally charged,  the surfacic charge in the fluid  
 $\sigma \vert\partial_{\scriptscriptstyle W}\Lambda\vert$ is also equal to zero (whether the dielectric constants  in the wall and in the medium where the fluid is embedded are equal or not).  
Indeed, in  the grand canonical 
 ensemble \eqref{defuXi},  the summation over microscopic states involves non-neutral
 configurations, but the self-energies of these globally charged configurations
 give them  exponentially vanishing weights in the thermodynamic limit, because they are not compensated by interaction energies with external charges inside the walls. 

Since the  bulk charge neutrality \eqref{neutrbulk} is satisfied for any set of chemical potentials,
the bulk densities $\roab$'s are determined by  only  $n_s-1$ independent functions of the $n_s$ chemical potentials $\mua$'s.
In the present paragraph we investigate more precisely the nature of the corresponding degeneracy.

By definition, the electrostatic  energy in the Hamiltonian 
\eqref{Hquant} used in $\Xi$ 
\eqref{defuXi} is the difference between the electrostatic energy of the interacting system and the energy 
$V_{\scriptscriptstyle R}\soma \ea N_{\alpha}$ of the noninteracting system in the reservoir where the electrostatic potential takes the uniform value $V_{\scriptscriptstyle R}$. In other words, the dependence of chemical potentials with respect to the potential 
$V_{\scriptscriptstyle R}$ in the reservoir is just
\begin{equation}
\label{muaVR}
\mua(V_{\scriptscriptstyle R})=\mua(0)+\ea V_{\scriptscriptstyle R}. 
\end{equation}

The global volumic and surfacic neutralities are linked to the invariance of  the thermodynamic limits of observables under a translation of the origin  for the electrostatic potentials. 
Indeed, if the latter origin is translated by an amount $-\Delta V$, then,
the reference  potential 
$V_{\scriptscriptstyle R}$ of the reservoir   becomes $V_{\scriptscriptstyle R}+\Delta V$,  the Hamiltonian is unchanged (since the insulating wall carries no external charge), and  the only change in $\Xi$ 
\eqref{defuXi} is an extra contribution $\Delta V\soma \ea 
 N_{\alpha}$ arising from $\soma \mua(V_{\scriptscriptstyle R}) N_{\alpha}$. Then the thermodynamic limit of $\ln \Xi$ is increased by 
\begin{equation}
 \Delta \left(\lim_{\rm th}\ln \Xi\right)=\beta  \Delta V
\,\lim_{\rm th}\langle\sum_{\alpha} \ea N_{\alpha}\rangle
\end{equation} 
According to \eqref{chargemacro}, \eqref{neutrbulk}, and \eqref{neutrsurf}, the latter variation vanishes up to order 
$\vert\partial_{\scriptscriptstyle W}\Lambda\vert$ included.

 Since  a translation $-\Delta V$ of the origin for the electrostatic potentials is  equivalent to an increase $\ea \Delta V$ of every  $\mua$,
 the nature of the degeneracy of physical quantities  with respect to chemical potentials is that physical quantities are invariant under the addition of an energy $\ea \Delta V$ to every chemical potential $\mua$. The 
 corresponding  degeneracy with respect to  fugacities $\za$'s  comes from the definition 
\begin{equation}
  \label{zneutrq}
 \za(V_{\scriptscriptstyle R}) \equiv 
\frac{ (2S_{\alpha}+1)}{\left(2\pi\laa^2\right)^{3/2}}
\exp\left[\beta\mua(V_{\scriptscriptstyle R})\right],
\end{equation}
The dependence of fugacities 
 upon $V_{\scriptscriptstyle R}$ is given by  \eqref{muaVR}.
The system involves charges of both signs, so that the 
continuous function
$f(V_{\scriptscriptstyle R}) = \soma \ea \za (V_{\scriptscriptstyle R})$  varies from $-\infty$ up to
$+\infty$ when $V_{\scriptscriptstyle R}$ varies from $-\infty$ to
$+\infty$. Therefore there exists a value of $V_{\scriptscriptstyle R}$ which fulfills the condition
$f(V_{\scriptscriptstyle R})=0$.


As a consequence, since physical quantities are invariant under a translation of 
$V_{\scriptscriptstyle R}$ in the fugacities, we can choose a set of fugacities which ensures that
the local charge neutrality in the bulk is enforced even in the zero-coupling limit, namely 
we can arbitrarily split the degeneracy with respect to fugacities 
 by imposing
\begin{equation}
  \label{zneutr} 
 \sum_{\alpha} \ea \za = 0.
\end{equation}
We notice that, as shown in Ref.\cite{Cornu04I}, in the case of an insulating wall with an external charge or in the case of a conducting wall, which becomes charged by influence, the global neutrality of the full system (the fluid plus the wall) in the thermodynamic limit implies that condition \eqref{zneutr} can also be fulfilled when $\Xi$ is written with the full Hamiltonian.
The ``neutrality'' condition about fugacities \eqref{zneutr} 
will cause major simplifications in the following calculations.


\subsection{Feynman-Kac formula}
\label{section23}

The non-commutativity between the kinetic and interaction operators in 
\eqref{defXi} 
can
be circumvented by using  Feynman-Kac formula \cite{Kac59,Simon}. The 
quantum Gibbs factor can be rewritten in terms of path integrals,
\begin{multline}
  \label{FK}
 \langle \{ \vecr_{i} \}
  \vert  e^{- \beta  \widehat{H}_{\{N_\alpha\}}}
   \vert \{ \vecr_i \}\rangle
= \left[ \prod_i \frac{1}{\left( 2 \pi \lambda_{\ai}^2 \right)^{3/2}}\right]
\int \left[\prod_i \Dvxii\right] \\
\times\exp \left[ - \beta \sum_{i<j} e_{\ai} e_{\aj} \int_0^1 ds \, \vb \left(
\vecr_i + \li\vxi_i (s)- \vecr_j - \lj\vxi_j (s) \right) \right]. 
\end{multline}
The kinetic part of the Hamiltonian and the particle-wall interaction
are taken into account in the measure $\Dvxii$ of the closed Brownian path 
$\vxi_i$ with dimensionless abscissa $s$: 
$\vxi (s=0) = \vxi (s=1) = \bf{0}$. The random path $\li \vxi_i (s)$ with typical extent $\li$ describes 
the quantum position fluctuations of particle $i$ at position $\vecr_i$. 
As discussed in Section \ref{section25}, the interaction between paths on the r.h.s. of \eqref{FK} is not the usual Coulomb
interaction between charged wires, since it involves only path elements with the 
same abscissa $s$.


The repulsion from the wall causes the anisotropy of the 
Brownian-path measure. The constraint about the quantum particle position, which  is described by $\Vsr(x)$,
enforces that the $x$-component $\xi_x$ of  vector $\vxi$ obeys the inequality 
\begin{equation}
    \label{condxix}
    x + \laa \xi_x (s) > 0 
  \end{equation}
for every $s$ between $0$ and $1$. 
The Brownian path measure can be factorized as
\begin{equation}
\label{measuretot}
\Dvxi= \Dxix\,{\cal{D}}(\vxi_{\|}),
\end{equation}
where $\vxi_{\|}$ is the projection of $\vxi$ onto the wall. 
As in the bulk,  the Gaussian measure ${\cal{D}}( \vxi_{\|})$ is independent of 
the position $\vecr$. Moreover it is rotational-invariant
and normalized to unity
\begin{equation}
  \label{mesxipar}
  \int  {\cal{D}}(\vxi_{\|})= 1.
\end{equation}
On the contrary, as recalled in Ref. \cite{jn&Fran99}, the measure $\Dxix$
depends on $x$, with for instance
\begin{equation}
  \label{mesqx}
\int \Dxix = 1 - e^{- 2 x^2/\laa^2}. 
\end{equation}
Moreover, the mean extent of the path in the $x$-direction does not vanish
\begin{equation}
\int\Dxix\,\, \xi_x \neq 0.
\end{equation}
All moments of the measure tend gaussianly fast to their bulk values over
the scale of the de Broglie wavelengths. For instance
\begin{equation}
  \label{intmesqx}
\int_0^1 ds \, \int \Dxix \, \, \xi_x (s) = \sqrt{\frac{\pi}{2}} \left(
\frac{x}{\lambda_\alpha} \right)^2 {\rm Erfc} \left( \sqrt{2} 
\frac{x}{\lambda_\alpha} \right),
\end{equation}
where ${\rm Erfc}(u)$ is the complementary error function defined as
\begin{equation}
{\rm Erfc}(u)=\frac{2}{\sqrt{\pi}} \, \int_u^{\infty} dt \, e^{-t^2}.
\end{equation}
${\rm Erfc}(u)$ decays as $\exp[-u^2]/(u\sqrt{\pi}) $ when $u$ goes to $+\infty$.


\subsection{Equivalence with a classical gas of loops}
\label{section25}
 
In the present paragraph we recall that the quantum grand partition function $\Xi$ for point particles can be written as a classical grand partition function $\Xi_{{\rm loop}}$ for randomly shaped loops \cite{Cornu96I}. 
The latter formalism 
including quantum statistics allows one to retrieve  property 
\eqref{relXiXiMB} : quantum statistics effects appear only at order
$(\lambda/a)^3$ in the low-degeneracy regime. 
In other words,  results at leading order in the low-degeneracy regime \eqref{lsa}
are the same when the starting partition function is written with  
Maxwell-Boltzmann statistics. Therefore,
for the sake of simplicity, 
    we shall directly consider $\Xi_{\rm MB}$  and we shall drop the index ${\rm MB}$ from now on.

The quantum grand partition function \eqref{defXi} can be rewritten by use of
the Feynman-Kac formula \eqref{FK} as 
 \begin{multline}
  \label{Xiqd}
  \Xi(\beta, \{\za\}, \Lambda) = \Xi_{{\rm loop}} \equiv
  \sum_{N=0}^{\infty} \frac{1}{N !} \int 
 \left[ \prod_{n=1}^{N}  d {\cal{L}}_n \,z({\cal{L}}_n ) \right]  \\ \times
\exp\left[- \beta \sum_{i<j} e_{\ai} e_{\aj} {\cal{V}}({\cal{L}}_i,
 {\cal{L}}_j)\right].
\end{multline}
In \eqref{Xiqd} the notation ${\cal{L}}\equiv ({\bf r}, \vxi, \alpha)$ 
stands for the loop position $\vecr$, 
the loop shape $\vxi$ and the loop species $\alpha$. 
When the measure is defined as 
\begin{equation}
\int d{\cal{L}}\equiv  \sum_{\alpha=1}^{n_s} \int_{\Lambda} d{\bf r}
\int \Dvxi,
\end{equation}
simple combinatorics allows one to replace the summation over the 
$N_\alpha$'s by a single summation over $N  = \soma N_\alpha$. 
The loop fugacity depends on the distance from the wall as 
\begin{equation}
  \label{defzL}
  z({\cal{L}}) = \za\,\theta(x),
\end{equation}
where $\theta (x) $ is the unit Heaviside function. 
The interaction between loops arising from the Feynman-Kac formula 
couples only line elements with the 
same abscissa $s$
\begin{equation}
  \label{defvbq}
 {\cal{V}}( {\cal{L}}_i, {\cal{L}}_j ) \equiv 
 \int_0^1 ds \, \vb \left(\vecr_i +
   \lambda_{\ai} \vxi_i(s)- \vecr_j - \lambda_{\aj} \vxi_j (s) \right).
\end{equation}
 Thus it is different from the electrostatic potential 
$ {\cal{V}}_{\rm elect}( {\cal{L}}_i, {\cal{L}}_j )$ between uniformly charged wires where any line element of a loop interacts with 
every line element of the other loop,
\begin{equation}
  \label{defvelect}
 {\cal{V}}_{\rm elect}( {\cal{L}}_i, {\cal{L}}_j ) \equiv 
 \int_0^1 ds  \int_0^1 ds' \, \vb \left(\vecr_i +
   \lambda_{\ai} \vxi_i(s)- \vecr_j - \lambda_{\aj} \vxi_j (s') \right).
\end{equation}


For a system with quantum statistics, $\Xi_{{\rm loop}}$ has the general expression written  in \eqref{Xiqd} where 
loops ${\cal{L}}$ and their fugacities 
$z({\cal{L}})$ have more complex expressions than in the case of Maxwell-Boltzmann statistics \cite{Cornu96I}. Quantum statistics is taken into account thanks to an extra internal degree of freedom, the number of particles  exchanged in the same permutation cycle.


Equality \eqref{Xiqd} between  the grand partition function of a quantum gas
of point particles and 
the grand partition function of a
classical system of loops with random shapes is the root of an
 equivalence between both systems.
As derived in Ref.\cite{Cornu96I}, the quantum density $\roa (x)$ can be determined from the loop density 
$\rho (\Lr)$ defined as a grand-canonical average calculated with 
$\Xi_{{\rm loop}}$. When exchange effects are neglected,  
\begin{equation}
\label{defrhoL}
\rho( \Lr ) \equiv \left< \sum_n \delta(\vecr_n - \vecr) \, 
\delta (\vxi_n - \vxi ) \, \delta_{\alpha_n,\alpha}
\right>_{\Xi_{{\rm loop}}},
\end{equation}
and  the relation between particle and loop
densities reads
\begin{equation} 
\label{corrho}
\roa(x) = \int \Dvxi  \, \,  \rho ( \Lr ),
\end{equation}
where $\rho (\Lr)$, with $\Lr = (\vecr, \vxi, \alpha) $,  does not depend on the projection $\vy$ of $\vecr$ 
onto the wall plane. 


\subsection{Ideal gas}
\label{section24}

In the case of an ideal quantum gas, 
the grand partition function \eqref{Xiqd}  is reduced to 
\begin{equation} 
\Xi^{\rm id}= \exp\left\{
\soma \za\int_{\Lambda}
d\vecr\int\Dvxi
\right\} +
\cO\left(\left(\frac{\laa}{a_{\alpha}}\right)^3\right),
\end{equation}
and, by virtue of \eqref{defrhoL}, $\rho^{\rm id}({\cal{L}})=z({\cal{L}})$ given in \eqref{defzL}.
The density
profiles of an ideal gas with fugacities $\za$'s are given by \eqref{relrhoXi} (or equivalently by \eqref{corrho}), and 
by using    \eqref{measuretot}--\eqref{mesqx} we retrieve that  
\begin{equation}
\label{defroaids} 
\rho^{{\rm id}}_{\alpha}(x) =\za
\left[1 - e^{- 2 x^2/\laa^2}\right]
+\za \cO\left(\left(\frac{\laa}{a_{\alpha}}\right)^3\right).
\end{equation}

We stress that, in the vicinity of the wall, the quantum charge fluid cannot be handled with as a system made of independent charges, because it cannot simultaneously obey the volumic global neutrality \eqref{neutrbulk} and the surfacic global neutrality \eqref{neutrsurf}. Indeed, the bulk densities in the ideal gas are equal to the $\za$'s.
If the constraint \eqref{zneutr} is arbitrarily enforced upon fugacities,  though there is no degeneracy with respect to fugacities  in the purely noninteracting case, the
  bulk densities satisfy the bulk local neutrality relation \eqref{neutrbulk}. However, the global surface charge  of the corresponding ideal gas does not vanish when species have different masses: 
according to 
\eqref{zneutr} and \eqref{defroaids}, it is equal to
\begin{equation}
\label{neutrsurfid}
\sigma^{{\rm id}} \equiv \int_0^{\infty} dx \, \soma \ea \rho^{{\rm id}}_{\alpha}(x)=
-\frac{1}{2}\sqrt{\frac{\pi}{2}} \soma \ea\za \laa. 
\end{equation} 



\section{Diagrammatic representation}

\label{diag}

The main lines of the following diagrammatic expansions are analogous to the
formalisms devised  for bulk quantum properties and classical density profiles 
in Refs. \cite{Cornu96I} and \cite{jn&fr01II}, respectively.

\subsection{Generalized fugacity-expansions}
\label{section31}

The equivalence with the classical loop system allows one to use 
techniques originally introduced for classical fluids. 
For instance, the Mayer diagrammatics initially built for point particles 
can be generalized to objects with internal degrees of freedom, such as
the species $\alpha$ or the loop shape $\vxi$. A generalized Mayer 
diagram for the loop density $\rho({\cal L})$  contains one root point $\Lr$ which is not integrated over and 
$N$ internal points ($N = 1, \ldots \infty$) which are integrated over, while
each pair of points is linked by at most one bond 
\begin{equation}
  \label{qdeff}
  f({\cal{L}}_i, {\cal{L}}_j) = 
  e^{- \beta \ei \ej  \calV ( {\cal{L}}_i , {\cal{L}}_j) } - 1.
\end{equation}
We choose to write the loop-fugacity expansion of the loop density as
\begin{equation}
  \label{expmayrq}
  \rho({\cal{L}}) = z({\cal{L}})
   \exp \left\{ \sum_{\cg} \frac{1}{S_{\cg}} \int \left[ \prod_{n=1}^N
  d{\cal{L}}_n z( {\cal{L}}_n ) \right] \left[ \prod f \right]_{\cg} \right\}.
\end{equation}
The summation is performed over all unlabeled, topologically different, connected
diagrams $\cg$ where the root point $\Lr$ is not an articulation point. An articulation point is defined by the following property: if it is taken out of the diagram, the latter 
is split into at least two pieces not linked together by any bond. (In another diagrammatic
representation \cite{Cornu96I}, which is analogous to \eqref{expmayrq} but without the exponential, the root point $\Lr$ may be an articulation point.)
$\left[ \prod f \right]_{\cg} $ 
is the product of the $f$-bonds in  diagram $\cg$  and $S_{\cg}$ is the 
symmetry factor, i.e. the number of permutations of  internal points 
$\Lr_n$ that do not change this product.

\begin{figure}[ht]
\begin{equation*}
\rho (\Lr ) \hspace{-1mm} = \hspace{-1mm} z (\Lr) \exp \left\{ 
\Lr \diagf \hspace{-0.5mm} z  + \Lr \diagdf  +  \Lr \diagtf  + \cdots 
\right\}
\end{equation*}
\caption{Mayer diagrammatic representation of the loop density. A white point stands 
for the root loop of the diagram, the coordinates of which are 
not integrated over. Black points are internal loop-points, 
the coordinates of which are integrated over. The $f$-bond between two
loops is represented by a line and the weight of a black point associated with a loop 
$\Lr_i$ is the loop fugacity $z(\Lr_i)$.}
\label{rhoMayer}
\end{figure}

At large distances with respect to de Broglie thermal wavelengths, the 
loop potential $\calV (\Lr, \Lr')$ behaves as the Coulomb potential 
between the total charges of loops, as if they were concentrated at 
positions $\vecr$ and $\vecr'$. Because of the latter $1/|\vecr - \vecr'|$ 
interactions, the integrals associated with generic diagrams $\cg$ in \eqref{expmayrq}
diverge in the thermodynamical limit.


\subsection{Systematic resummations of large-distance  \\ Coulomb divergences}
\label{section32}

The large-distance divergences arising from the long-range of  Coulomb potential can be dealt with by introducing auxiliary
bonds and by systematically resumming subclasses of auxiliary diagrams $\tilde{\cg}$. The
method is a generalization of the procedure introduced by Meeron for bulk quantities in a classical
Coulomb fluid \cite{Meer58}.
The systematic resummation procedure is displayed in Appendix \ref{appresum}
 where  similarities and differences with the process used in Ref.\cite{Cornu96I} are  stressed.

As in Ref.\cite{Cornu96I},
the decomposition into 
auxiliary bonds relies on the multipolar decomposition of the loop interaction. This decomposition  allows
 one to exhibit classical screening  through the appearance of a screened potential $\phi$ arising from the resummation process.
It reads
\begin{equation}
  \label{decompvwq}
  {\cal {V}}(\Lr_i, \Lr_j) = {\cal {V}}^{\rm cc}( \vecr_i - \vecr_j ) 
  + \vcm ( \vecr_i, \Lr_j ) +
  \vmc(\Lr_i, \vecr_j) + \vmm (\Lr_i, \Lr_j ).
\end{equation}
where ${\cal {V}}^{\rm cc}( \vecr_i - \vecr_j ) =v(\vecr_i - \vecr_j)$ is the charge-charge -- i.e. monopole-monopole -- interaction between the total loop
charges,
and the charge-multipole interaction $\vcm$ is equal to
\begin{equation}
  \label{vcmi}
  \vcm (\vecr_i, \Lr_j) \equiv \int_0^1 ds \, 
  \Bigl[ v \left( \vecr_i -  \vecr_j -
  \lj \vxi_j (s) \right) - v \left( \vecr_i - \vecr_j \right) \Bigr],
\end{equation}
with a symmetric definition for $\vmc$. The multipole-multipole interaction
$\vmm$ is
\begin{multline}
  \label{vmmi}
  \vmm ( \Lr_i,\Lr_j)  = {\cal{V}} ( \Lr_i, \Lr_j) - v ( \vecr_i - \vecr_j) - \vcm
  ( \vecr_i, \Lr_j) - \vmc ( \Lr_i, \vecr_j)   \\
 = \int_0^1 ds \left[ \rule{0mm}{5mm} 
 \vb \left( \vecr_i + \lambda_i \vxi_i (s) - \vecr_j -
  \lambda_j \vxi_j(s) \right) - \vb \left( \vecr_i + \lambda_i \vxi_i (s)
  -  \vecr_j  \right)  \right.  \\
 \left. \rule{0mm}{5mm}  - \vb \left( \vecr_i - \vecr_j -
  \lambda_j \vxi_j(s) \right) + \vb \left( \vecr_i - \vecr_j \right) \right].
\end{multline}
$\vcm$ does coincide with the charge-multipole Coulomb interaction between a
point charge and a charged wire, whereas $\vmm$ is not equal to the
multipole-multipole Coulomb interaction between two charged wires, because it couples only  line elements of $\Lr$ and $\Lr'$ with the same abscissa. 
The resummation procedure introduces a screened potential $\phi$ arising from the sum of all 
chains built with the auxiliary bond \linebreak 
$\fcc (\Lr, \Lr') = - \beta \ea \eap v(\vecr - \vecr')$ (see Fig.\ref{figphi}). It reads
\begin{multline}
  \label{defchain}
   - \beta e_\alpha e_{\alpha'} \, \phi ( \vecr, \vecr') = 
\fcc (\Lr , \Lr') \\   
 +  \sum_{N=1}^{\infty} \int \left[ \prod_{n=1}^N 
 d\Lr_n \, z (\Lr_n) \right] \fcc (\Lr,
    \Lr_1) \, \fcc (\Lr_1, \Lr_2) \ldots \fcc (\Lr_N, \Lr').
\end{multline}
The properties of $\phi$ are studied hereafter.

\begin{figure}[ht]
\begin{multline}
\nonumber
  - \beta e_{\alpha} e_{\alpha'} \, \phi ( \vecr, \vecr') 
= \Lr \diagbFcc \Lr' \\ 
= \Lr \diagfcc \Lr' + \Lr \diagdfcc \Lr' + \Lr \diagtfcc \Lr' + \ldots
\end{multline}
\caption{Screened potential  $\phi$.}
\label{figphi}
\end{figure}

When resummed bonds are defined, associated resummed weights and ``{excluded-composition}'' rules 
ensure a one-to-one correspondence between each class in the partition of auxiliary diagrams $\tilde{\cg}$ 
and each resummed diagram $\cpps$. 
Contrary to what is done in Ref.\cite{Cornu96I}, we choose to consider nine resummed bonds defined hereafter. The reason is that
the choice of these nine bonds is associated with  renormalized
 weights  which are more convenient for dealing with the case of a wall with  an electrostatic response (see Section \ref{section7}) than the renormalized weights 
 which appear when only the five resummed bonds of Ref.\cite{Cornu96I} are retained.

\begin{figure}[ht]
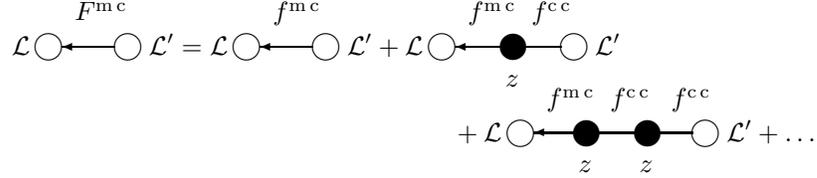

\begin{multline}
\nonumber
 \Lr  \diagbFmc \Lr'
 = \Lr \diagfmc \Lr' + \Lr \diagfmcfcc \Lr' \\
+ \Lr \diagfmcdfcc \Lr' + \ldots
\end{multline}
\caption{Resummed bond $\Fmc$.}
\label{figFmc}
\end{figure}

\begin{figure}[ht]
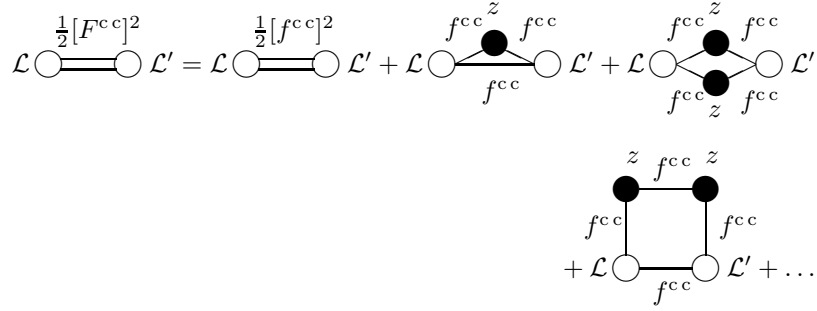

\begin{multline}
\nonumber
 \Lr  \diagbFccd \Lr'
 = \Lr \diagbfccd \Lr' + \Lr \diagfccdfcc \Lr'  + \Lr \diagdfccdfcc \Lr'  \\ 
\rule{0mm}{1.3cm}\\
+ \Lr \diagfcctfcc \Lr'  + \ldots
\end{multline}
\caption{Resummed bond $\frac{1}{2} \left[ \Fcc \right]^2$. (The symmetry factors are not written in the figure.)}
\label{figFccd}
\end{figure}

The nine  resummed bonds are 
\begin{subequations}
\label{defF}
  \begin{equation}
    \label{defFcc}
    \Fcc ( \Lr, \Lr') = - \beta \ea \eap \, \phi (\vecr, \vecr'),
  \end{equation}
\begin{equation}
    \label{defFmc}
    \Fmc (\Lr, \Lr') = - \beta \ea \eap \int_0^1 ds \, \left[ \phi (\vecr +
    \laa \vxi (s), \vecr' ) - \phi (\vecr, \vecr') \rule{0mm}{4mm} \right]
\end{equation}
(see Fig.\ref{figFmc}) with a symmetric definition for $\Fcm$,   
$(1/2)\left[\Fcc\right]^2$ (see Fig.\ref{figFccd}), $(1/2) \left[\Fcm \right]^2 $,
 $(1/2) \left[\Fmc \right]^2$, $\Fcc. \Fcm$, $\Fcc. \Fmc$, and  
\begin{multline}
    \label{defFr}
    \Frt ( \Lr, \Lr') =  \left\{ \rule{0mm}{5mm} e^{\Fcc+ \Fcm+\Fmc+\Fmm} - 1 - 
    \Fcc - \Fcm - \Fmc \right. \\ 
 \left.  - \frac{1}{2} \left[\Fcc \right]^2 - \frac{1}{2} \left[\Fcm \right]^2 -
 \frac{1}{2} \left[\Fmc \right]^2 - \Fcc. \Fcm - \Fcc. \Fmc  \right\} (\Lr, \Lr').
  \end{multline}
\end{subequations}
As shown in Ref.\cite{Cornu96I}, 
\begin{equation}
\label{valueFmm}
\Fmm( \Lr, \Lr') = - \beta \ea \eap \, 
\phi^{\rm mm} ( \Lr, \Lr')+W( \Lr, \Lr'), 
\end{equation}
where $\phi^{\rm mm}$ is defined as $\vmm$ \eqref{vmmi} with $\phi(\vecr,\vecr')$ in place of $v(\vecr-\vecr')$, and
$W$ is a purely quantum contribution,
\begin{eqnarray}
\label{defW}
W( \Lr, \Lr')&=&
- \beta \ea \eap \,\left[{\cal{V}}(  \Lr, \Lr' )-
{\cal{V}}_{\rm elect}(  \Lr, \Lr' )\right]\\
&=&- \beta \ea \eap \, \int_0^1 ds  \int_0^1 ds'
\left[\delta(s-s')-1\right]
\sum_{q=1}^{+\infty}\sum_{q'=1}^{+\infty}
\frac{1}{q!}\frac{1}{q'!}\nonumber\\
&&\qquad\qquad\times
\left[\lambda_{\alpha} \vxi (s) . \nabla_{\vecr} \right]^q 
\left[\lambda_{\alpha'} \vxi' (s') . \nabla_{\vecr'} \right]^{q'}
v(\vecr-\vecr').\nonumber
\end{eqnarray}
We notice that the argument of the exponential in \eqref{defFr} can be written $\Fcc+ \Fcm+\Fmc+\Fmm=F_{\rm elect}+W$, where 
\linebreak $F_{\rm elect}= 
- \beta \ea \eap \int_0^1 ds \int_0^1 ds'\,  \phi (\vecr +
    \laa \vxi (s), \vecr' + \laap \vxi' (s'))$.

Eventually, the Mayer fugacity-expansion \eqref{expmayrq} of the loop density can be rewritten as
\begin{equation}
  \label{expmayrqf}
  \rho(\Lr) = \zs(\Lr)
   \exp \left\{ \sum_{\cpps} \frac{1}{S_{\cpps}} \int \left[ \prod_{j=1}^N
  d\Lr_j \, w( \Lr_j ) \right] \left[ \prod F \right]_{\cpps} \right\}.
\end{equation}
The effective screened fugacity arising from the resummation of Coulomb 
ring diagrams (defined in Appendix \ref{appresum} and  Fig.\ref{Ir}) is
equal to 
\begin{equation}
\label{defzs}
\zs(\Lr) =  z(\Lr) \, e^{-\beta \ea^{2} \calVs(\Lr)},
\end{equation}
where $z(\Lr)=\theta(x) \za$ and 
\begin{equation}
\label{defVs}
\calVs(\Lr) = \frac{1}{2} \int_0^1 ds \int_0^1 ds'\,
\left[ \phi - \vb \right] \left( \vecr + \laa
  \vxi (s) , \vecr + \laa \vxi (s') \right).
\end{equation} 
The $\cpps$ diagrams are defined as the $\cg$ diagrams in \eqref{expmayrq}
apart from the following two differences. First the $f$-bond is replaced by the nine $F$-bonds. 
Second, $\cpps$ diagrams obey an ``{excluded-composition}'' rule associated
with the fact that all points have not the same weight $w(\Lr)$ (in order to  avoid double-counting),
\begin{equation}
\label{poidsres}
w(\Lr)=\left\{ \begin{array}{ll}
                   \zs(\Lr)- z(\Lr) &\mbox{ if $\Lr$ is involved only in a  } \\
                            &   \mbox{
product $F^{{\rm a \, c}} (\Lr_i, \Lr) \, F^{{\rm c \, b}} (\Lr,\Lr_j)$}\\
                   \zs(\Lr)  &\mbox{ otherwise.}
                            \end{array}
                            \right.
\end{equation}
In \eqref{poidsres}
  superscripts $a$ and $b$ stand either for 
${\rm c}$ or ${\rm m}$, and the points $\Lr_i$ and  $\Lr_j$ may coincide.
\begin{figure}[ht]
\begin{minipage}[b]{0.96\linewidth}
\epsfig{figure=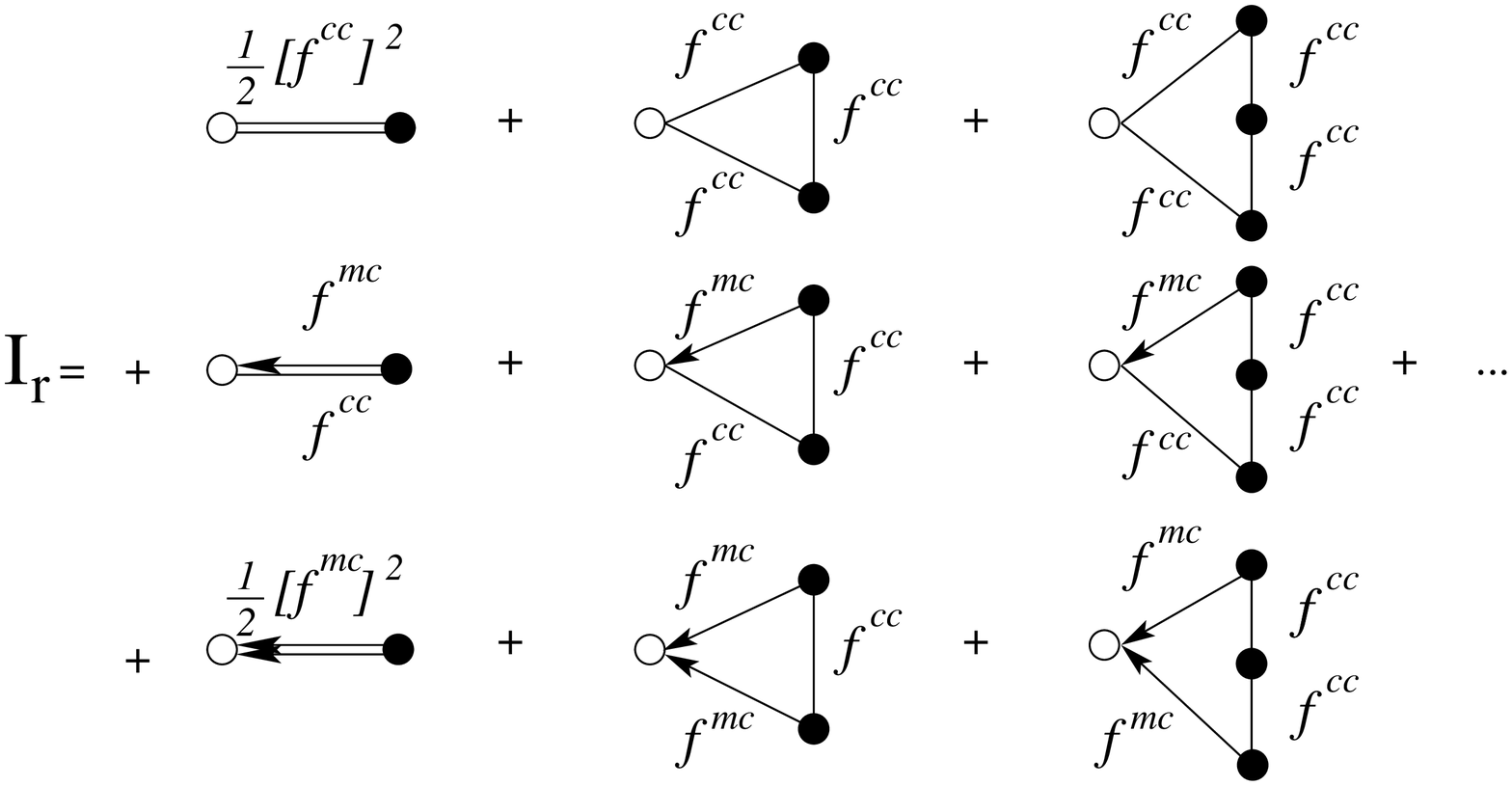,width=10.5cm}
\caption{The sum of ring diagrams attached to a white point, $\Ir = - \beta \ead \calVs$.}
\label{Ir}
\end{minipage}
\end{figure}

The vicinity of the wall replaces the bulk exponential screening by an integrable algebraic   screening in the directions parallel to the wall (see next section), and  all diagrams with resummed bonds are finite in the thermodynamical limit, when integrations are performed first over loop shapes and then over loop positions. The reasons are the following ones. First, in $\Fcc$ \eqref{defFcc} the resummed  charge-charge interaction  between the total loop charges is proportional to the screened potential $\phi(\vecr,\vecr')$ which obeys an inhomogeneous Debye equation where the screening length depends on the distance from the wall, as discussed in next section. The translational invariance along the wall  ensures that  
$\phi(\vecr,\vecr') =\phi(x,x',\vy)$ where $\vy$ is the projection of 
$\vecr - \vecr'$ onto the wall. 
 As shown in an analogous classical situation \cite{AquaCor03I},  where the inhomogeneity in the zero-coupling limit arises from the electrostatic response of the wall,
 when $\vert\vy\vert$ goes to infinity 
while the values of $x$ and $x'$ are kept fixed
$\phi(\vecr,\vecr')$ decays as
\begin{equation}
\phi(\vecr,\vecr')\mathop{\sim}_{\vert\vy\vert\rightarrow +\infty}
\frac{f(x,x')}{\vert\vy\vert^3}.
\end{equation} 
Far away from the wall the $x$-dependent screening length tends to a non-zero value, so that   
$f(x,x')$ decays exponentially fast to zero at large  $x$ or  $x'$. Therefore the leading tail of $\Fcc$ has the structure 
\linebreak
$g^{\rm cc}(x, x')/\vert\vy\vert^3$ and $\Fcc$ is integrable.  In $\Fmc$ \eqref{defFmc} the 
resummed \linebreak
multipole-charge interaction is also integrable. Indeed, since the Brownian path measure $\Dvxi$ ensures that all moments of $\vxi$ are finite, and 
the leading tail in $\Fmc$, 
which is not canceled by the integration over the loop shape $\vxi$, behaves as 
\begin{equation}
\label{tailFmc}
\int_0^1 ds\, \sum_{q=1}^{+\infty}\frac{1}{q!}
\left[\lambda_{\alpha} \xi_{x} (s)\right]^q
\frac{\partial^q g^{\rm cc}(x,x')}{\partial x^q}
\times \frac{1}{\vert\vy\vert^3}.
\end{equation} 
$\partial^q g^{\rm cc}(x,x')/\partial x^q$ is  an exponentially vanishing  function of $x$ and $x'$ far away from the wall, while $\int_0^1 ds\,\Dvxi \left[\xi_x(s)\right]^q$  is a gaussianly vanishing function of $x$ over the scale $\lambda_{\alpha}$. As a consequence, the bond $\Fcm$ is also integrable in $x$ and $x'$ over the scales $\lambda_{\alpha}$ and $\lambda_{\alpha'}$, respectively. By virtue of \eqref{defFr}
 the tail of $\Frt ( \Lr, \Lr')$ at large distances is given by the tail of the resummed multipole-multipole interaction in 
$\Fmm$ \eqref{valueFmm} and can be decomposed into a ``{diffraction}'' term  arising from $\phi^{\rm mm} $ and a purely quantum contribution described by $W$ defined  in \eqref{defW}. The corresponding leading  tails are
\begin{equation}
\label{tailFmm}
\int_0^1 ds \int_0^1 ds'\, 
\sum_{q=1}^{+\infty}\sum_{q'=1}^{+\infty}
\frac{1}{q!}\frac{1}{q'!}
\left[\lambda_{\alpha} \xi_{x} (s)\right]^q
\left[\lambda_{\alpha'} \xi'_{x'} (s')\right]^{q'}
\frac{\partial^{q+q'} g^{\rm cc}(x,x')}{\partial x^q\partial x'^{q'}}
\times \frac{1}{\vert\vy\vert^3}
\end{equation} 
 and
\begin{equation}
\label{tailW}
-\beta \ea \eap \int_0^1 ds  \int_0^1 ds'\left[\delta(s-s')-1\right]
\lambda_{\alpha} \xi_{x} (s)
\lambda_{\alpha'} \xi'_{x'} (s')
\frac{\partial^2 v(\vecr-\vecr')}{\partial x\partial x'}
\end{equation}
The tail \eqref{tailFmm}  is integrable for the same reason as the tail \eqref{tailFmc} of $\Fmc$.
When $\vert\vy\vert$ goes to infinity 
while the values of $x$ and $x'$ are kept fixed, the  function \eqref{tailW} decays as $\vert\vy\vert^3$ times a function
which, after integration over the Brownian measures $\Dvxi $ and $\Dvxip$, depends on  $x$ and $x'$ and converges gaussianly fast to zero over the scale $\lambda_{\alpha}$ and 
$\lambda_{\alpha'}$ respectively. Eventually,
thanks to resummations, every diagram in  \eqref{expmayrqf} is 
well defined in the thermodynamical limit, and we will consider this 
limit from now on.

\section{Screened potential} 
\label{ldlim}

\subsection{Debye equation for an inhomogeneous fluid} 
\label{section33}

The screened potential $\phi$ which is defined as the sum of chains 
\eqref{defchain} is the solution of the integral equation 
\begin{equation}
\label{eqintphi}
 \phi(\vecr,\vecr')=v(\vecr,\vecr') 
-\frac{1}{4\pi} \int  d{\vecr}''\, \kbar^2 (x'') 
\, v(\vecr,\vecr'') \, \phi(\vecr'',\vecr'). 
\end{equation}
In \eqref{eqintphi} the positive function  $\kbar^2(x)$ reads
\begin{eqnarray}
  \kbar ^2 (x)  &\equiv& \, 4 \pi \beta \sum_{\alpha}\ea^2 \int \Dvxi \, 
  z(\Lr)\nonumber \\
   &=&  \, \theta(x) \, \, 
 4 \pi \beta \sum_{\alpha} \ea^2 \za \left[ 1 - 
e^{- 2 x^2 / \laa^2} \right], 
\end{eqnarray}
where the second equality arises from \eqref{measuretot}--\eqref{mesqx}.
Far away from the wall $\kbar^2(x)$ tends towards its bulk value 
\begin{equation}
  \label{}
  \kz^2 \equiv 4 \pi \beta \soma \ea^2 \za. 
\end{equation}

By using  Poisson equation satisfied by Coulomb potential $v$,
\begin{equation}
\Delta_{\vecr} v(\vecr, \vecr') 
   = - 4  \pi
  \delta( \vecr - \vecr'),  
\end{equation}  
the integral 
equation \eqref{eqintphi} is shown to be equivalent to a set of partial 
derivative equations, the explicit expressions of which depend on the signs of $x$ and $x'$. For $x'>0$, $\phi(\vecr, \vecr')$ is a solution of 
\begin{subequations}
\label{Debyeinhom}
\begin{equation}
\label{eqdifphi}
   \Delta_{\vecr} \phi (\vecr, \vecr') - \kbar^2 (x) \, \phi(\vecr, \vecr')  
   = - 4  \pi
  \delta( \vecr - \vecr')  \qquad \mbox{for}  \quad x >0 
\end{equation}
and
\begin{equation}
   \Delta_{\vecr} \phi ( \vecr, \vecr') = 0  \qquad \mbox{for}  \quad x < 0.
\end{equation}
\end{subequations}
Moreover, the diagrammatic definition \eqref{defchain} of $\phi$ implies that $\phi$ obeys the same 
boundary conditions as the electrostatic potential $v(\vecr,\vecr')$,
\begin{subequations} 
\label{eqch}
\begin{eqnarray}
  & & \phi ( \vecr, \vecr') \quad \mbox{and} \quad \left.\frac{\partial \phi(\vecr, \vecr')}{\partial
  x} \right\vert_{x\not=x'} \quad \mbox {are continuous at $x=0$}  \label{condflimq1}\\
 & &  \lim_{|\vecr| \rightarrow \infty }\phi(\vecr, \vecr') =0. 
 \label{condflimq2}
\end{eqnarray}
\end{subequations}


The invariance of the system in directions parallel to the interface 
implies that the Fourier transform of $\phi$ along these directions obeys a
one-dimensional differential 
equation with respect to $x$.  We introduce the dimensionless
coordinates  $\xt \equiv \kz x$, $\vyt \equiv \kz\vy$ and $\vecrt \equiv \kz \vecr$
and the dimensionless screened potential $\phit$ defined by
\begin{equation}
  \label{defphit}
\phit (\xt,\xt',\vyt)=\frac{1}{\kz}\phi(x,x',\vy).  
\end{equation}
The dimensionless Fourier transform of $\phit$ along directions parallel to the
wall reads
\begin{equation}
  \label{deftf}
  \phit(\xt,\xt',\vq) =  \int d\vyt \, e^{-i \vq \cdot \vyt}
 \phit (\xt, \xt',\vyt).
\end{equation}
For $\xt'>0$ it is a solution of the differential equations
\begin{subequations}
\label{equadiffq}
\begin{equation}
  \label{equadiffuq}
  \left\{ \frac{\partial^2}{\partial \xt^2} - \left( 1 + \vq^2 \right) 
- U(\xt) \right\} \phit (\xt, \xt', \vq)= - 4 \pi \delta ( \xt - \xt')
\quad \mbox{ for $\xt>0$}
\end{equation}
and
\begin{equation}
  \label{equadiffu2}
  \left\{ \frac{\partial^2}{\partial \xt^2} -  \vq^2  \right\} 
  \phit (\xt, \xt', \vq)= 0
\quad \mbox{ for $\xt<0$,}
\end{equation}
\end{subequations}
with the potential function
\begin{equation}
  \label{potUq}
  U (\xt) \equiv 
 -  \frac{\soma \ea^2 \za e^{- 2 \xt^2/(\kz \laa)^2}}{\somg \eg^2 \zg}.
\end{equation}


\subsection{Series representation of the exact screened potential}
\label{section34}

It is well known (see e.g. Ref. \cite{Zwillinger}) that  
the solution of the one-dimensional differential equation
\eqref{equadiffuq} 
can be formally written in terms of the solutions $h$ of the 
associated ``{homogeneous}'' equation (where
the Dirac distribution on the r.h.s.  is replaced by zero and) which is valid for $-\infty<\xt<+\infty$,
\begin{equation}
  \label{equadiffhom}
  \left\{ \frac{\partial^2}{\partial \xt^2} - \left( 1 + \vq^2 \right) 
- U(\xt) \right\}h (\xt ; \vq^2)= 0.
\end{equation}
More precisely, a  solution $\phit$ of \eqref{equadiffuq} can be decomposed 
into the following sum:  a linear combination of two independent 
solutions $h^+$ and $h^-$ plus   a particular solution 
$\phit_{\rm sing}$ which is singular when $x=x'$ and is 
calculated in terms of  $h^+$ and $h^-$ by the so-called Wronskian
method.  
In the following, $h^+$ ($h^-$) is chosen to be  a solution which  vanishes 
(diverges) when $\xt$ tends to $+\infty$. Then $\phit_{\rm sing}$ reads
\begin{equation}
  \label{phitbI}
  \phit_{\rm sing} (\xt, \xt',\vq^2) = - \frac{4 \pi}{W(\vq^2)} \,
   h^- \Bigl( \inf(\xt,\xt'); \vq^2 \Bigr)  
h^+  \Bigl( \sup(\xt,\xt'); \vq^2 \Bigr),  
\end{equation}
where the Wronskian $W(\vq^2) \equiv h^- (dh^+/d\xt) - h^+ (dh^- / d\xt)$ 
is independent of $\xt$. 
Since the differential operator in \eqref{equadiffuq} 
is self-adjoint, the real potential $\phit$ has the symmetry
$\phit(\xt,\xt',\vq)=\phit(\xt',\xt,\vq)$ when  $\xt>0$ and $\xt'>0$. Then
the boundary condition \eqref{condflimq2} 
obeyed by $\phit$ enforces that
\begin{equation}
\label{structphi}
\phit(\xt,\xt',\vq)= \phit_{\rm sing}(\xt,\xt',\vq^2) +Z(|\vq|) \, 
h^{+} (\xt; \vq^2) \, h^{+} (\xt'; \vq^2),
\end{equation}
where $Z(|\vq|)$ is determined by the continuity relations \eqref{condflimq1}.

 The particular solution $h^+$ ($h^-$) of  the
``{homogeneous}'' equation \eqref{equadiffhom} can be written as 
the solution $\exp[-\xt\rac]$ ($\exp[\xt\rac]$) of \eqref{equadiffhom}
when $U=0$, times $1+H^+$ ($1+H^-$) where the function $H^+$  ($H^-$), which describes boundary effects, is chosen to vanish at $\xt=0$, 
\begin{equation}
\label{tutu}
h^{\pm}(\xt;\vq^2)=e^{\mp\xt\rac} [1+H^{\pm} (\xt, \vq^2)].
\end{equation}
Then
\begin{multline}
\label{marc}
\phit_{\rm sing}(\xt,\xt',\vq^2)=-\frac{4\pi}{W(\vq^2)} 
e^{- \vert\xt-\xt'\vert \rac}\\
\times[1+H^-(\inf(\xt,\xt');\vq^2)][1+H^+(\sup(\xt,\xt');\vq^2)].
\end{multline}
The Wronskian $W(\vq^2)$ proves to take the simple form
\begin{equation}
W(\vq^2)= -2\rac+\left.\frac{\partial H^+(\xt;\vq^2)}{\partial \xt}
\right\vert_{\xt=0}.
\end{equation}
Indeed, $H^+(\xt;\vq^2)$ and $H^-(\xt;\vq^2)$ vanish 
at $\xt=0$,  and so does   $\partial H^-/\partial \xt$,
as can be checked  on the formal solution given in next paragraph.

As shown in Ref.\cite{jn&fr01II},
$H^+$ can be written as the formal alternative series 
\begin{equation}
  \label{hpserie}
 H^+(\xt;\vq^2) = - {\cal T}^{+}[1] (\xt;\vq^2) + {\cal T}^{+} 
  \left[{\cal T}^+ [1] \right] (\xt;\vq^2) \rule{0mm}{5mm}
- \cdots  
\end{equation}
where the operator ${\cal T}^+$ acting on a function $f$ reads
\begin{equation}
  \label{defcalTp}
  {\cal T}^{+} [f] (\xt;\vq^2) \equiv  \int_0^{\xt} dv \, e^{2 v \rac} 
 \int_v^{+\infty} dt \, e^{-2 t \rac}  \, U (t)  f(t).
\end{equation}
The solution $H^-$ can be
similarly written in terms of another formal series
\begin{equation}
  \label{hmserie}
 H^{-} (\xt;\vq^2) = {\cal T}^{-}[1] (\xt;\vq^2) + {\cal T}^{-} 
  \left[{\cal T}^- [1] \right] (\xt;\vq^2) \rule{0mm}{5mm}
+ \cdots  
\end{equation}
with
\begin{equation}
\label{defcalTm}
  {\cal T}^- [f] (\xt;\vq^2) \equiv  \int_{0}^{\xt} dv \, e^{-2 v \rac} 
 \int_0^{v} dt \, e^{2 t \rac}  \, U(t)  f(t).
\end{equation}
 By combining \eqref{structphi}--\eqref{defcalTm},  we get a representation of the screened potential $\phit$ 
in terms of  formal series. 

These series are bounded by geometric series of $\kzl$
(with $\lambda=\sup_{\alpha}\{\lambda_{\alpha}\}$), because 
 the Gaussian function $U$ is integrable for all $x$'s.
More precisely,  a straightforward calculation shows that 
\begin{multline}
  {\cal T}^+ [1] (\xt) = - \frac{\sqrt{\pi}}{4 \sqrt{2} \rac} 
\frac{\soma \ead \za \klat}{\somg \eg^2 \zg} \left\{
 \Erf \left( \frac{\sqrt{2} \, \xt}{\klat} \right)\right. \\ 
+e^{\left( \klat \rac \right)^2 /2} 
 \left[ e^{2 \xt \rac} \, 
\Erfc \left( \frac{\sqrt{2} \, \xt}{\klat} + 
              \frac{\klat \rac}{\sqrt{2}} 
            \right)\right. \\
 \left.\left.-\Erfc \left( \frac{\klat \rac}{\sqrt{2}} \right) 
\right]  
\right\},
\end{multline}
where $\klat=\kz \lambda_{\alpha}$, and this expression obeys the inequality 
\begin{equation}
  \label{boundT1}
  \left|  {\cal T}^+ [1] (\xt) \right| <\kzl.
\end{equation}



\subsection{$\kzl$-expansion of the screened potential}
\label{section42}

In Section \ref{weak}
we will  restrict our explicit calculations to a low-degeneracy 
 and weak-coupling regime
(see  \eqref{lsa} and \eqref{petitG}). In this regime  
$\kzl $ is also negligible with respect to $1$ : the de Broglie 
thermal wavelengths $\laa$'s are small compared with the typical screening length 
$\kz^{-1}$.


Let $\phitzero $ be the solution of equation \eqref{equadiffuq} when the profile ${\kbar}^2(x)$ is replaced by its bulk limit $\kz^2$. 
The above formal series for $H^{+}$ and $H^{-}$ 
provide a systematic expansion of  $\phit-\phitzero $  in terms of the ratios of the length
scales $\laa$'s, over which $\kbar ^2(x)$ varies, and the length scale 
$\kz^{-1}$, which is the 
limit of ${\kbar}^{-1} (x)$ when $x$ goes to infinity. 
The expansions of $H^{+}$ and $H^{-}$  
  are absolutely and uniformly convergent 
with respect to the variable $x$ for values of $\kzl$ smaller than some finite value, because  they are bounded by geometric series. As a result,
\begin{equation}
  \label{bob}
  \phit ( \xt, \xt', \widetilde{\bf y}) =  
 \phitzero( \xt, \xt', \widetilde{\bf y}) + \cO ( \kz \lambda), 
\end{equation}
where the leading-order term $\phitzero$ is equal to 
 $\kz^{-1} \phizero (\vecr, \vecr')$ (see \eqref{defphit}). By definition,
$\phizero(\vecr, \vecr') $ is the
solution of
\begin{equation}
\label{eqphiz}
 \Delta_{\vecr} \phizero (\vecr, \vecr') - \theta (x)  \, \kz^2 \,  
 \phizero(\vecr, \vecr')  
   = - 4  \pi  \delta( \vecr - \vecr'),
\end{equation}
with the same boundary conditions as $\phi (\vecr, \vecr')$. 
According to \eqref{structphi}, we get
\begin{equation}
\label{resufiz}
\phitzero(\xt,\xt',\vq)=\frac{2 \pi}{\rac} \left\{ 
e^{- \vert \xt-\xt' \vert \rac}
+ \frac{\rac - q}{\rac + q} e^{- (\xt+\xt')\rac} \right\}.
\end{equation}

Moreover, by  definition of  the screened potential, 
the correction $\cO ( \kz \lambda) $ in \eqref{bob} is 
independent of the root-point species $\alpha$ and $\alpha'$. As 
shown in Section \ref{section52}, the latter property implies that
  the contribution to density profiles from
 this  $\cO ( \kz \lambda) $-correction in $\phit$ is  canceled at first order in $\kz \lambda$ by the neutrality constraint 
\eqref{zneutr} on fugacities.  It  is the reason why  we do not write the explicit expression of the term $\cO ( \kz \lambda) $ in \eqref{bob}.



\section{Weak-coupling expansions}

\label{weak}

 \subsection{Subregime for parameters $\eps$ and $(\lambda/a)^3$}
\label{section51}

In the weak-coupling and  low-degeneracy  regime \eqref{petitG} and \eqref{lsa}, only a finite number of resummed Mayer diagrams in the representation
 \eqref{expmayrqf} of the loop-density $\rho(\Lr)$ contribute
 at lowest orders in the classical coupling parameter $\eps\equiv (1/2)\kz \beta e^2$.
Moreover diagram contributions may  be also expanded in powers of $\kzl$. 

Indeed,
$\kzl$ is a function of the two independent parameters $(\lambda/a)^3$ and $\eps$: by virtue of \eqref{rels},
\begin{equation}
\label{relkzleps}
\kzl\propto \frac{\lambda}{a}\,\eps^{1/3},
\end{equation} 
and $\kzl\ll 1$ in the considered regime 
 \eqref{petitG} and \eqref{lsa}.
Then the screened potential $\phi$, which is involved both in screened fugacities and in resummed bonds, can be expanded in powers of 
$\kzl$, as well as the functions resulting from  integrations over Brownian paths.
As shown by the scaling analysis performed in next Sections
\ref{section43} and \ref{section52}, 
the diagrammatic representation  of the particle density derived from \eqref{corrho} and \eqref{expmayrqf}  provides a
systematic expansion of $\roa(x)$ in powers of  parameters $\eps$ and $\kzl$, where exchange effects are neglected. 

More precisely, we will show that, because of quantum dynamics, the first coupling corrections to the ideal-gas particle density is a sum of two terms of order $\eps$ and $\kzl$ respectively,
and the next
 contributions are of order 
\begin{equation}
\label{defetad}
\eps^2,\,  \eps^2\vert\ln(\kzl)\vert,\, \eps \, . \, \kzl,\, (\kzl)^2,\,
 \left( \frac{\lambda}{a} \right)^3  
Q_{\scriptscriptstyle W}\left(-\frac{\beta e^2}{\lambda}\right), 
 \end{equation}
with
\begin{equation}
\label{ordcoupl}
 \left( \frac{\lambda}{a} \right)^3 \propto
\frac{ \left(\kzl\right)^3}{\eps}\quad{\text {and}}\quad
\frac{\beta e^2}{\lambda}\propto\frac{\eps}{\kzl}
\end{equation} 
 $Q_{\scriptscriptstyle W}(t)$ is expected to vanish at $t=0$ by analogy with the bulk function $Q$ defined in Ref.\cite{Ebeling69}.
On the other hand, 
exchange corrections, which  have been neglected from the start,  are of order 
\begin{equation}
\label{ordquant}
\left( \frac{\lambda}{a} \right)^3 E_{\scriptscriptstyle W}\left(-\frac{\beta e^2}{\lambda}\right)
\end{equation} 
 where $E_{\scriptscriptstyle W}(t)$ is expected to vanish as  $t$ goes to zero, for the same reason as $Q_{\scriptscriptstyle W}(t)$.

As a consequence, we shall be allowed to retain only the corrections linear in $\eps$ and $\kzl$ in $\roa(x)$, if these terms are larger than the exchange corrections of order \eqref{ordquant} and  the coupling corrections of order \eqref{ordcoupl}.
We consider  the subregime where 
\begin{equation}
\label{exchnegeps}
\eps^2 \leq \left(\frac{\lambda}{a}\right)^3\ll \eps,
\end{equation}
In \eqref{exchnegeps} $\eps^2 \leq \left(\lambda/a\right)^3$ means that   the ratio of $\eps^2$ and $\left(\lambda/a\right)^3$ is either kept fixed or tends to zero when both $\eps$ and $\lambda/a$ vanish. In other words, $\beta e^2/\lambda$ is kept fixed -- i.e. the  temperature remains fixed -- or vanishes
-- i.e. the temperature goes to infinity.  In this subregime the condition
\begin{equation}
\left( \frac{\lambda}{a} \right)^3  
F\left(-\frac{\beta e^2}{\lambda}\right)\ll \eps,
\end{equation} 
with $F=Q_{\scriptscriptstyle W}$ or $E_{\scriptscriptstyle W}$, is met, as well as the condition 
\begin{equation}
\left( \frac{\lambda}{a} \right)^3  
F\left(-\beta e^2/\lambda\right)\ll \kzl.
\end{equation} 
Similarly,  inequalities 
$ \eps^2\ll \kzl$, $\eps^2\vert \ln(\kzl)\vert\ll \eps$,  $\eps^2\vert\ln(\kzl)\vert\ll \kzl$,
and $ (\kzl)^2\ll \eps$, are also satisfied. 
Eventually,  in the subregime \eqref{exchnegeps}, which 
can be rewritten as 
$\eps^3\leq (\kzl)^3\ll \eps^2$ by virtue of \eqref{relkzleps}, we may  retain only contributions of order $\eps$ and $\kzl$, and the neglected terms are of order
$\cO \left( \eta^2 \right)$, where $\eta^2$ is a generic notation for the terms in \eqref{defetad} and \eqref{ordquant},
\begin{equation}
\label{defeta}
\cO \left( \eta^2 \right)\equiv\cO(\eqref{defetad},\eqref{ordquant}).
\end{equation}


\subsection{Screened loop fugacity }
\label{section43}

In this section we calculate the $\kz \lambda$-expansion of the screened 
fugacities \eqref{defzs}. As shown in Ref. \cite{jn&fr01I},
the free energy $\ead \calVs$ is associated with the ``{geometric}''
repulsion from the  wall due to the deformation of the screening cloud surrounding every 
charge near a boundary. According to \eqref{defVs} and the
fact that $(\phi - v)$ and its derivative are continuous
(because $\phi$ and $v$  have the same singularity when $\vecr=\vecr'$), 
\begin{equation}
\label{expcalVs}
\calVs(\Lr) = \Vs(\vecr) +\frac{1}{\kz}\cO(\kz\lambda).
\end{equation}
\eqref{expcalVs} is the Taylor expansion of $\calVs$ around its classical value
\begin{equation}
\Vs(\vecr)\equiv \frac{1}{2} \left[ \phi - \vb \right] ( \vecr  , \vecr ).
\end{equation}
(We stress that $\Vs(\vecr)$ has no singularity in the range $0\leq x$ so that no classical spurious singularity is introduced by the expansion  \eqref{expcalVs}.)
The $\kz \lambda$-expansion \eqref{bob} of the screened potential $\phit$ leads to
\begin{equation}
\Vs(\vecr)= \Vszero( \vecr) +\frac{1}{\kz}\cO(\kz\lambda),
\end{equation}
where at first order
\begin{equation}
\label{M}
-\beta \ead\Vszero(\vecr)=
 -\frac{\beta \ead}{2}\left[ \phizero - \vb \right] ( \vecr  , \vecr )
 =\epsa \left[1-\Lb (\kz x)\right].
\end{equation}
In \eqref{M} we have used the definition
\begin{equation}
  \label{defepsa}
  \epsa \equiv \frac{1}{2} \kz \beta \ead. 
\end{equation} 
By virtue of \eqref{resufiz}, 
\begin{eqnarray}
  \label{defLbq}
  \Lb ( u) 
&\equiv& \int_1^{\infty} dt \, \frac{e^{- 2 t u
  }}{\left( t + \sqrt{t^2 - 1 } \right)^2 } \nonumber\\
 &=& e^{- 2 u } \left[ 
 \frac{1}{2u} + \frac{1}{u^2} + \frac{1}{2 u^3 }
  \right] - \frac{1}{u} K_2 ( 2 u),
\end{eqnarray}
where $K_2 (2u)$ is a Bessel function, which decays proportionally to \linebreak
$\exp(-2u)/\sqrt{u}$ at large $u$.
$\Lb (u)$ is a continuous positive decreasing function for $u \geq 0$. Therefore 
$\Lb ( u) $ is  bounded 
by $\Lb(0)=1/3$ and, since $\epsa$ is small, we can expand the exponential function in the 
definition \eqref{defzs} of screened loop fugacities. We get
\begin{equation}
\label{dimanche}
 z^{\rm sc}(\Lr)
=\theta(x) \, \za \left\{ \rule{0mm}{5mm}
1 + \epsa \left[1-\Lb(\kz x)\right] +\cO(\eps \, . \, \kzl)
\right\}.
\end{equation}
Since $\Lb (u)$ vanishes exponentially fast 
when $u$ goes to infinity,  $z^{\rm sc}(\Lr)$ tends to
$\za \left\{ 1 + \epsa \right\} $ far away from the wall.


\subsection{Diagram contributions at leading orders}
\label{section52}

\subsubsection{Bond $\Fcc$}

The integral associated with the diagram with one $\Fcc$-bond   reads
\begin{equation}
  \label{eqcu}
\int d\Lr' z^{\rm sc}(\Lr') \Fcc(\Lr,\Lr')= - \beta \ea \somg \eg \int d\vecr' \phi (\vecr, \vecr') 
\,\zbscg (x'),
\end{equation}
where the screened fugacity $\zbsca (x)$ is defined as
\begin{equation}
\label{defzscpart}
\zbsca (x)\equiv \int \Dvxi \, z^{\rm sc}(\Lr).
\end{equation}
By using the value \eqref{dimanche} of the screened loop fugacity and the 
integral \eqref{mesqx} of the path measure, we get
\begin{equation}
\label{valuezscreen}
\zbsca (x)
=\za \left[1-e^{-2{x}^2/\laa^2}\right]
\left\{ \rule{0mm}{5mm} 1+\epsa\left[1-\Lb(\kz x)\right]\right\} +\za\cO(\eps.\kzl).
\end{equation}
By virtue of  \eqref{bob}, the leading-order terms in \eqref{eqcu} arise from the structure
\begin{equation}
  \label{eqcd}
  - \beta \ea \somg \eg \zg \int_{x'>0} d\vecr' \,
 \left[\phi^{(0)}( \vecr,\vecr') + \cO ( \kz \lambda) \right]
    \left[1 - e^{-2 {x'}^2 / \lag^2} \right] \Bigl[ 1 + \cO_\gamma 
(\eps) \Bigr],
\end{equation}
where $\cO_\gamma(\eps)$ is a term of order $\eps$ which depends on the
species $\gamma$. ($\cO_\gamma(\eps)$ is the screened self-energy term in  $\zbscg (x)$.)

The \textit{a priori} leading-order term in \eqref{eqcd} is 
obtained  by retaining  only the two constants $1$ in 
brackets and $ \phi^{(0)}( \vecr,\vecr')$.
This term is independent of the de Broglie wavelengths, because
$ \phi^{(0)}(\vecr, \vecr')$ is purely classical and  involves only the length scale $1/\kd$. By virtue of 
\eqref{resufiz},
\begin{equation}
\label{intyphizero}
\int_{x'>0} d\vecr' \, \phi^{(0)}(\vecr, \vecr')
=\frac{2\pi}{\kz^2} \int_0^{+\infty} d\xt'
\left\{e^{-|\xt-\xt'|}+e^{-(\xt+\xt')}
\right\}.
\end{equation} 
As a consequence, the leading term in \eqref{eqcd} is both
  $\cO(\eps^0)$ and $\cO\left((\kzl)^0\right)$, namely,
before summation over $\gamma$,
\begin{equation}
\label{prop0}
\int d\vecr' \int \Dvxigp \, z^{\rm sc}(\Lr') \Fcc(\Lr,\Lr') =\cO(1).
\end{equation}

However, after summation over species,  the  $\cO(1)$
 contribution in \eqref{eqcu}, where
the quantity $\somg \eg \zg$ is factorized,   is exactly equal to zero because of the neutrality condition \eqref{zneutr} imposed on fugacities.   Moreover, without  condition
\eqref{zneutr},  
we would have to consider an infinite number of diagrams at leading order, because the addition of a ``{star}'' 
subdiagram 
$\prod_{i=2}^{n}\left[\int d\Lr_i \zs ( \Lr_i) 
\Fcc(\Lr',\Lr_i)\right]$, with an arbitrary number $n$, to the diagram $\int d\Lr' \zs ( \Lr') \Fcc(\Lr,\Lr')$ would also yield a contribution of leading order $\cO(1)$.
Nevertheless, we notice that, after summation over all diagrams, the final  expansion of the density must be 
independent of whether the condition on fugacities is fulfilled or not, 
because of the degeneracy among fugacities discussed in Section \ref{section22}.

In fact, the
diagram with one bond $\Fcc$ contributes at orders $\eps$ and $\kzl$,
\begin{equation}
\label{orderdiagFcc}
\int d\Lr' z^{\rm sc}(\Lr') \Fcc(\Lr,\Lr')=\cO(\eps,\kzl)
\end{equation} 
The contribution of 
order $\cO(\eps)$ comes from $\phizero (\vecr, \vecr') \, \times \, \cO_\gamma (\eps)$
in \eqref{eqcd}, and it has been calculated in Ref. \cite{jn&fr01II}.
The contribution of order $\kzl$ arising from 
the product of constants $1$ in \eqref{eqcd} times the term of order $\kzl$ in the expansion of $\phi(\vecr,\vecr')$ is canceled by the neutrality condition \eqref{zneutr}. (This is also true for the whole $\kzl$-expansion of 
$\phi(\vecr,\vecr')$, because
all terms which depend on a species $\gamma$ only through the product $\eg\zg$ are canceled when the summation over $\gamma$ is performed.) 
Therefore,  
the contribution  of  order $\kzl$ arises only from
$\phizero (\vecr, \vecr') \exp [ - 2 {x'}^2 / \lag^2 ]$
in \eqref{eqcd}.  
The  factor $\kzl$ is yielded by the
$x'$-integration  
of the Gaussian term which arises from the integrated quantum measure \eqref{mesqx}.
Indeed,  for any bounded and integrable function $f$ which decays less quickly than $\exp [ - 2 {x}^2 / \laa^2 ]$ at large $x$
\begin{eqnarray}
\label{prop2bis}
 \int_0^{+\infty} \kz dx \, \, e^{-2x^2/\laa^2} f(\kz x) 
\mathop{\sim}_{\kz \laa \rightarrow 0}&&
\,\kz \laa \times f(0) \int_0^{+\infty}dt\, e^{-2t^2}\\
= &&
\cO \left( \kz \laa\right) \times
 \cO \left(\int_0^{+\infty} du \,\, f(u)\right).\nonumber
\end{eqnarray}
(The Gaussian factor in the integral on the l.h.s. of \eqref{prop2bis} makes this
integral convergent over the scale $\lag$, and not $\kz^{-1}$ 
as it would be the case if this factor  were not here.)
 As shown in Appendix \ref{appscal}, the same mechanism  also operates for
the  $x'$-integration of  odd moments of a Brownian path $\vxi'$ (involved in
diagrams  
with $\Fcm$-bonds), because these moments decay gaussianly fast  
to zero at large $x'$ over the scales $\lambda_{\alpha}$'s (see for instance  \eqref{intmesqx}), contrary to even moments which tend to their non-zero bulk values away from the wall.

\subsubsection{Other resummed bonds }

 As shown in Appendix \ref{appscal}, by combining previous arguments,
Taylor expansions around classical expressions   and construction rule \eqref{poidsres}
about fugacities, where $\zs (\Lr') - z(\Lr')  = \za \cO(\eps )$,  a simple scaling analysis  shows that diagrams with one bond,
$\Fcm$, $\Fmc$,  $[\Fcc]^2/2$,
$\Fcc\Fcm$, $[\Fcm]^2/2$, $\Fcc\Fmc$, or $[\Fmc]^2/2$  are  of orders equal to either $(\kzl)^2$, $\eps\cdot\kzl$, $\eps^2$,
$\eps(\kzl)^2$,  $\eps^2\cdot\kzl$, or $\eps^2\cdot(\kzl)^2$.
The order of the contribution from the bond $\Frt$ is determined by analogy with its order in the bulk case.

\subsubsection{Global results} 

As a result, the complete scaling analysis shows that, at first order in  $\eps$ and $\kzl$, the loop density comes from \eqref{expmayrqf} 
where only one diagram, namely the diagram with a single $\Fcc$-bond, is retained in the argument of the exponential. 
Moreover, the 
 expression of this first-order contribution involves 
only the lowest-order term $\phizero (\vecr, \vecr')$ in the 
$\kzl$-expansion \eqref{bob} of the screened potential. 
 The loop density  reads
\begin{multline}
\label{coucou}
\rho(\Lr)=z^{\rm sc}(\Lr) \\
\times \exp \left\{
-\beta\ea\sum_{\gamma} \eg\int d\vecr' \phizero(\vecr,\vecr') \left(\int
\Dvxigp \, z^{\rm sc}(\Lr') \right) \right.
\\
\left. \rule{0mm}{6mm} +\cO \left( \eqref{defetad} \right) \right\}.
\end{multline}
\begin{figure}[ht]
\begin{equation*}
\rho (\Lr ) = \zsc (\Lr) \exp \left\{ \Lr 
\diagFcc z^{{\rm sc}} (\Lr') + \cO ( \eta^2) \right\}
\end{equation*}
\caption{Diagrammatic expression of the loop density at first order in
$\eps$ and $\kzl$.}
\label{figrhoou}
\end{figure}

Since the argument of the exponential in \eqref{coucou} proves to be a bounded function of order $\eps$ and $\kzl$, the loop density is given  at first order in $\eps$ and $\kzl$ 
by linearizing the exponential in \eqref{coucou}. 
According to \eqref{corrho}, the particle density profile is obtained by performing  the path integration $\int
\Dvxig$ with the result   
\begin{equation}
\label{rostruc}
\roa(x)=\zbsca (x) \left\{1
-\beta\ea\sum_{\gamma} \eg\int d\vecr' \, \zbscg(x') \, 
\phizero(\vecr,\vecr')  +\cO(\eta^2)\right\},
\end{equation}
where $\zbsca (x)$ is given in \eqref{valuezscreen} and $\eta^2$ is defined in \eqref{defeta}.


\subsection{Electrostatic potential}
\label{section53}

In this section we show how the integral involved in the expression \eqref{rostruc} 
is related to the  potential $\Phi(x)$, which is defined as the difference between the electrostatic potential created by the fluid and the electrostatic potential $V_{\scriptscriptstyle R}$ in the particle reservoir (located in the bulk). This potential 
obeys  Poisson equation
\begin{equation}
\label{intforPhi}
\frac{d^2 \Phi}{d x^2} (x)=-4\pi \soma\ea \roa(x)
\end{equation}
with the boundary conditions : $\Phi$ and $d \Phi / dx$ tend to
$0$ when $x$ goes to $+ \infty$. The condition about the derivative of $\Phi$ arises from
the absence of any  net electrostatic field in the bulk for a Coulomb fluid at 
equilibrium. The condition about $\Phi$ fixes the electrostatic potential reference. 
The definition of $\Phi(x)$ leads to the integral representation
\begin{equation}
\label{intrelPhi}
\Phi(x)= - 4 \pi \int_x^{+\infty} dx' (x'-x) \soma \ea \roa (x').
\end{equation}

As in the classical case \cite{jn&fr01II}, 
the structure \eqref{rostruc} of density profiles can be rewritten in the form
\begin{equation}
\label{strucroG}
\roa(x) =\zbsca(x) \Bigl[  1-\beta \ea \, G(x)\Bigr]
\end{equation}
with
\begin{equation}
\label{defG}
G(x)\equiv \sum_{\gamma} \eg\int d\vecr' \, \zbscg(x') \, \phizero( \vecr,\vecr').
\end{equation}
As checked in Section \ref{section54}, at leading order $G(x)$ is a function of $\kz x$. Therefore, according to \eqref{valuezscreen} and \eqref{prop2bis}, the contribution from 
\linebreak
$-\beta \soma \ea^2 \left[\zbsca(x)-\za \right] G(x)$ to the integral in \eqref{intrelPhi} is a correction of relative orders
$\cO ( \kzl)$ and $\cO ( \eps)$ with respect to the leading contribution from
\begin{equation}
\label{dif}
 \soma \ea \zbsca(x)-\beta\left(\soma \ead \za\right) G(x).
\end{equation}
Therefore, the argument of Section 5.4 in Ref. \cite{jn&fr01II} holds. It reads as follows. According to its definition and the partial derivative 
equation \eqref{eqphiz} 
satisfied by $\phizero$, $G(x)$ obeys the differential equation
\begin{equation}
  \label{eqG}
\frac{d^2 G(x) }{d x^2} =  - 4 \pi \soma \ea \zbsca (x) + \kz^2 G (x).
\end{equation}
Since $dG(x)/dx $ is finite and decays 
to zero when $x$ goes to infinity, combination of \eqref{intrelPhi}, \eqref{dif}, and \eqref{eqG} implies that
 the electrostatic potential is merely
\begin{equation}
\label{calculPhi}
\Phi(x)=\left[ G(x) -\lim_{x\rightarrow +\infty}G(x) \right]+\frac{1}{\beta e}\cO(\eps,\kzl),
\end{equation} 
where $\cO(\eps,\kzl)$ denotes a sum of terms of order 
$\eps$ and $\kzl$ respectively.


Eventually, by virtue of \eqref{valuezscreen} and
\eqref{calculPhi}, the density profile \eqref{strucroG} can be 
rewritten in terms of the bulk density and the
electrostatic potential drop $\Phi(x)$ created by the fluid as
\begin{equation}
\label{exprfnphi}
\roa (x) =  \roab \, \left[1-e^{-2x^2/\laa^2}\right]
\left\{ \rule{0mm}{4mm} 1 - \epsa \Lb(\kz x) 
 - \beta \ea \Phi(x) \right\}+ \roab\cO(\eta^2),
\end{equation}
where the bulk density is given by
\begin{equation}
\label{resultroab}
\roab\equiv \lim_{x\rightarrow +\infty}\roa(x)
=\za
\left\{1+\epsa -\beta \ea \lim_{x\rightarrow +\infty}G(x) + \cO ( \eta^2)
 \right\}.
\end{equation}
Since the bulk density 
$\roab$ coincides with $\za$ at leading order,
$\kd$ defined in \eqref{defxid} is also equal to $\kz$ at leading order
\begin{equation}
  \label{relkdkz}
  \kd 
= \kz \Bigl[ 1 + \cO ( \eps) \Bigr] \quad\text{and}\quad
\epsd=\eps\Bigl[ 1 + \cO ( \eps) \Bigr].
\end{equation}
This will enable us to consider the Debye screening length as the reference length scale for 
classical effects when writing the final results in next section.



\subsection{Decomposition into classical and quantum contributions}
\label{section54}

In the density profile \eqref{exprfnphi}, the term involving $\Lb$ is purely classical, 
whereas the electrostatic potential can be split into a classical contribution $\Phiclu$ of order 
$\cO\left(\eps/(\beta e)\right)$ and a quantum contribution $\Phiquu$ of order $\cO\left(\kzl/(\beta e)\right)$,
\begin{equation}
\label{decompPhi}
\Phi(x)= \Phiquu(x)+ \Phiclu(x)+  \frac{1}{\beta e}\cO \left(
\eta^2 \right).
\end{equation}
According to \eqref{calculPhi}, 
 $\Phiclu (x) = G^{{\rm cl} \, (\eps)}(x) - \lim_{x \rightarrow 
+ \infty} G^{{\rm cl} \, (\eps)} (x)$ with, by virtue of \eqref{defG}, \eqref{valuezscreen}, and of  the neutrality 
constraint \eqref{zneutr} upon fugacities,
\begin{equation}
\label{Gcl}
G^{{\rm cl} \, (\eps)} (x)= \sum_{\gamma} \eg\int d\vecr' \, \theta(x') \, \zg
\epsg \left[1-\Lb(\kz x')\right]
\phizero(\vecr,\vecr'),
\end{equation}
whereas $\Phiquu$ is the leading term in the expansion of
\begin{equation}
\label{Gqu}
-\sum_{\gamma} \eg\int d\vecr' \, 
\theta(x') \, \zg \, e^{-2{x'}^2/\lag^2} \, \phizero(\vecr,\vecr'). 
\end{equation}
Since $\int d\vy \phizero(x,x',\vy)$ is a function of $\kz x'$ which decays exponentially fast when $\kz x'$ goes to infinity, according to \eqref{prop2bis},
\begin{equation}
\label{Phiquvalue}
\Phiquu (x)= \int_0^{+\infty} dx' \, \left(-\sum_{\gamma} \eg
z_{\gamma} \, e^{-2{x'}^2/\lag^2} \right)\,\int d\vy \phizero(x,x'=0,\vy). 
\end{equation}

The classical part  $\Phiclu$ in the electrostatic potential has already been calculated in Ref.{\cite{jn&fr01I}}  with the result 
\begin{equation}
\label{rappelPhicl}
\Phiclu (x)= - \Pucl \, \Mb (\kd x ), 
\end{equation}
where  the function $\Mb$ is  
\begin{equation}
  \label{}
  \Mb(u) = \int_1^{\infty}dt \, \frac{e^{-2 t u} - 2 t e^{-u}}{1-(2t)^2}
\frac{1}{\left( t + \ract \right)^2},
\end{equation}
and the constant $A$ reads
\begin{equation}
  \label{valueA}
  \Pucl = \sqrt{\pi \beta} \frac{\somg \eg^3 \rogb}{\sqrt{\soma \ead \roab}}.
\end{equation}
In \eqref{rappelPhicl},  the argument of 
$\Mb$ has been written  $\kd x$ in place of $\kz x$, by virtue of \eqref{relkdkz}.

  The purely quantum  part of the 
electrostatic potential drop is derived from \eqref{Phiquvalue}. 
According to the expression of $\int d\vy \phizero(x,x',\vy) $ already used in 
\eqref{intyphizero}, 
\begin{equation}
\label{valuePhiqubis}
\Phiquu (x)= - \hbar \, \B \, e^{-\kd x},  
\end{equation}
where the constant $\B$ depends only on the fluid composition 
\begin{equation}
  \label{valueB}
  \B = \frac{\pi}{\sqrt{2}} 
\frac{\somg (\eg/\sqrt{\mg}) \rogb}{\sqrt{\soma \ead \roab}}.
\end{equation}
We notice that the total electrostatic potential drop between the 
wall and the fluid bulk (set as the reference of electrostatic potentials) is equal at leading order to
\begin{equation}
  \label{valuePhi0}
  \Phi(0) = - \frac{\Pucl}{2} 
    \Bigl(\ln 3 - 1 - \frac{\pi}{\sqrt{3}}\Bigr) - \hbar \, \B. 
\end{equation}
$\Phi(0)$ includes both  classical and quantum corrections.

The structure of quantum  particle densities derived from 
\eqref{exprfnphi}, \eqref{rappelPhicl} and \eqref{valuePhiqubis} is summarized in 
\eqref{qdensfnch}. In the fugacity expansion of the bulk density $\roab$, 
the explicit first-order correction to the value $\za$ in an ideal gas is derived from \eqref{resultroab} and the explicit value of $G(x)$.
Since $G^{{\rm qu} (\kzl)} (x)$ tends towards  zero when $x$ goes 
to infinity, the relation between $\roab$ and $\za$ does not 
include quantum contributions proportional to $\hbar$ and reads
\begin{equation}
  \label{toto}
  \roab = \za \left\{ 1 + \beta \ead 
\sqrt{\pi \beta \somg \eg^2 \rogb}
- \beta \ea \sqrt{\pi \beta} \frac{\somg 
\eg^3 \rogb}{\sqrt{\soma \ea^2 \roab}}  + \cO(\eta^2) \right\}.
\end{equation}
 Indeed, in the bulk, the spherical symmetry enforces 
quantum-dynamical coupling
effects to be at least of order $(\kzl)^2$, 
proportional to $\hbar^2$, whereas exchange
effects are at least of order $\plsap^3$, proportional to $\hbar^3$. 
The expression \eqref{toto} is in agreement with the result (5.28) of Ref.\cite{ACP94} calculated directly in the  bulk. 
 It does satisfy the electroneutrality 
condition \eqref{neutrbulk}.


  The expression \eqref{qdensfnch} of the quantum density profile at first order in  $\eps$ and $\kz \lambda$ can be rewritten in terms of the 
densities $\roa^{\rm cl}$'s at first order in $\eps$
in the corresponding classical system,
\begin{equation}
\label{valuerhocl}
\roa^{{\rm cl}(\eps)}(x)=\roab \left[1-\epsa \Lb(\kd x) -\beta \ea \Phiclu (x) \right], 
\end{equation}
where $\Lb$ is defined in  \eqref{defLbq}. At this leading order the classical density profile does not involve the short-range repulsion that must be introduced in order to prevent the collapse of the system in the limit where $\hbar$ tends to zero
(see e.g. Ref.\cite{jn&fr01I}). Indeed, \eqref{valuerhocl} is obtained in a subregime where the range $\sigma$ of the short-distance repulsion is such that  $\eps^2\leq (\sigma/a)^3\ll \eps$. We get
\begin{equation}
\label{relrhoqurhocl}
\roa(x)=
\left[1-e^{-2x^2/\laa^2}\right]\left\{\rule{0mm}{5mm}
\roa^{{\rm cl}(\eps)}(x)- \roab\beta \ea \Phiquu (x)\right\}
+\roab \cO(\eta^2).
\end{equation}
The latter expression displays two quantum effects. We stress that the effect linked to the vanishing of wave-functions has  an essential singularity in $\hbar$. The quantum contribution linear in $\hbar$ in the electrostatic potential is allowed by the breakdown of spherical symmetry, whereas bulk quantum effects in the physical regime of interest appear only at order $\hbar^2$ \cite{ACP94}, as already mentioned.



\section{Generic properties}

\label{prop}

\subsection{Density profiles}

The structure \eqref{qdensfnch} of density profiles is ruled by 
the competition between three effects. The purely quantum contribution 
$\left[1-e^{-2 x^2 / \laa^2}\right]$ arises from  the vanishing of wave-functions
 inside the wall. 
In the physical regime of interest 
this effect is the same one as in an ideal gas (see \eqref{rhoid}), up to amplitude corrections arising from Coulomb coupling. 
The second term, involving the function $\Lb$, 
describes the geometric repulsion due to the deformation 
of screening clouds near a wall \cite{jn&fr01I}. Indeed, 
a charge and its surrounding screening cloud are more stable in a 
spherical geometry than in the dissymmetric configurations enforced by the presence 
of a wall. This effect is purely classical at the order of the present  calculation.
The term 
$\ea \Phi(x)$ describes the interaction between a particle with a charge $\ea$ 
and the electrostatic potential drop, 
created by the fluid itself, 
with respect to the bulk (set as the reference of electrostatic potentials). This contribution contains
both quantum and classical effects.

In the very vicinity of the wall, for $x \leq \lambda\ll \xid$ (with $\lambda=\sup_\alpha\{\laa\}$), densities are 
mostly ruled by the quantum  effect of the cancellation of wave-functions inside the wall,
\begin{equation}
  \label{compxlll}
  \roa (x) \underset{x \leq \lambda\ll\xid}{\sim} \roab \, \,
\left[1 - e^{- 2 x^2/\laa^2}\right]
 \left\{ 1 - \frac{1}{6} \kd \beta \ead
- \beta \ea \Phi(0) 
\right\},
\end{equation}
where $\Phi(0)$ is given in \eqref{valuePhi0}.
The heavier a particle species   is, the steeper the vanishing of its density occurs, since dynamical quantum effects are less
important for heavy particles. Densities and their first derivatives are continuous on the wall, as expected, since densities involve the squared moduli of wavefunctions and the latter ones are continuous at the boundary of a wall with a possible step variation in their first derivatives.

At distances  from the wall large compared with the quantum de Broglie 
wavelengths, $x \gg \lambda$, densities vary over the 
classical Debye screening length
\begin{equation}
  \label{rhointerm}
  \roa (x) \underset{ \lambda\ll x\leq \xid}{\sim} \roab \left\{ 1  - \frac{1}{2} \kd \beta \ead \Lb (\kd x)
+ \beta \ea
  \Bigl[  \Pucl \, \Mb (\kd x) 
+\hbar \, \B \, e^{-\kd x} \Bigr]
 \right\},
\end{equation}
where  $\Pucl$ and $\B$  are given in 
\eqref{valueA} and \eqref{valueB} respectively.
 In this 
region, density profiles are determined by the interplay 
between the effect of the classical geometric repulsion described by $\Lb (\kd x)$ and
the effect of the electrostatic potential, with  both 
 classical and quantum origins. 
$\Lb (\kd x)$ and $\Mb (\kd x)$ vanish exponentially fast over the scales 
$\xid/2$ and  $\xid$ respectively.

As a consequence, at distances from the wall large  compared with the classical
Debye screening length $\xid$, the contribution from  the electrostatic potential dominates in the density profiles
\begin{equation}
  \label{compasdens}
  \frac{\roa (x) - \roab}{\roab} \underset{x \gg \xid}{\sim} 
- \beta \ea  \Pas \, e^{- \kd x}, 
\end{equation}
where 
\begin{equation}
  \label{valuePhias}
  \Pas = 
- \frac{\Pucl}{8} \left[ \ln 3  + \frac{\pi}{\sqrt{3}}- 2 \right]
- \hbar \, \B. 
\end{equation}
Equation \eqref{compasdens} is valid in the generic case
where $\Pas \neq 0$. According to \eqref{valueA} and \eqref{valueB}, $\Pas$ is likely to vanish only in a two-component plasma where charges are opposite, $e_-=-e_+$ ($\Pucl=0$), and where  both species have the same mass
($\B=0$). When $\Pas \neq 0$,
if the charge $\ea$ has a sign opposite to that of $\Pas$,  
  $\roa(x) > \roab$ at sufficiently large distances $x$, as 
shown by \eqref{compasdens}.


\subsection{Profile of the total particle density}

At the order of calculations, the wall is  repulsive for the global 
particle density everywhere in the Coulomb  fluid,
as in the classical 
case \cite{jn&fr01I}, 
\begin{equation}
  \label{borndensitpart}
  \soma \roa (x) < \soma \roab.
\end{equation} 
Indeed, according to the structure  \eqref{qdensfnch} of densities, since $\roa(x)$ and 
\linebreak $1-\exp\left[-2x^2/\laa^2\right]$ are positive, the second factor  on the r.h.s. of \eqref{qdensfnch} is also positive in the considered regime of small parameters. Therefore the effect of the vanishing of wavefunctions near the wall is to lower the density $\roa(x)$:
\eqref{qdensfnch}  implies that at any distance $x$ from the wall
\begin{equation}
\label{borne}
\soma\roa(x)< \soma
\roab \left[ 1 
- \frac{1}{2} \kd \beta \ead \, \Lb (\kd x) 
-\beta\ea \Phi(x)\right].
\end{equation} 
The contribution from  the  electrostatic 
potential drop $\Phi(x)$ to the  bound in \eqref{borne} vanishes   because 
of the bulk electroneutrality \eqref{neutrbulk}, as in the case
of $\soma \roa^{\rm cl(\eps)}(x)$. Thus the  bound involves only the sum of the contributions from classical screened self-energies. The corresponding  geometric repulsion from the wall tends to reduce the density of each species with respect to its bulk value, and we get 
 \eqref{borndensitpart}.



\subsection{Charge density profile}

Even if the Coulomb fluid remains globally neutral,  
when species have different masses,
 the local 
charge density $\soma \ea\roa(x)$ is non zero, as well as the associated electrostatic potential drop $\Phi(x)$. The property holds even in the case
of a charge-symmetric two-component plasma  where the classical charge density
$\soma \ea\roa^{\rm cl}(x)$ vanishes
for symmetry reasons (because the two species have opposite charges). (The latter cancellation  can be checked  at first order in $\eps$ where, by virtue of \eqref{valuerhocl},
$ \soma \ea \roa^{{\rm cl}(\eps)}$ is proportional to  $\somg \eg^3 \rogb$.)
The charge density profile is organized in various layers with opposite signs which depend on the composition of the fluid.

In the very vicinity of the wall ($x \leq \lambda$),
when species have different masses,
the charge-density profile exhibits a zeroth-order effect arising from the cancellations of the various
wave-functions over different scales. Indeed, 
according to \eqref{compxlll} and \eqref{neutrbulk},
\begin{equation}
  \soma \ea \roa (x) \underset{x\leq\lambda\leq \xid}{\sim}
-\soma \ea \roab e^{-2x^2/\laa^2} + e\rho\,\cO\left( \eps,\kzl\right),
\end{equation} 
where $\rho$ is the typical particle density. When all masses are equal, \linebreak
$\soma \ea \roa (x)$ is only of order 
$e\rho\,\cO\left( \eps,\kzl\right)$ in this region, by virtue of the bulk local charge neutrality.

 At distances from the wall large with respect  
to the quantum lengths, $x \gg \lambda$, the charge density is an effect of order $e\rho\,\cO\left( \eps,\kzl\right)$ which is ruled by the competition between the geometric repulsion from the wall and the electrostatic potential 
drop (see \eqref{rhointerm}). According to \eqref{relrhoqurhocl} and \eqref{valuePhiqubis}, 
\begin{equation}
\soma \ea \roa (x)\underset{\lambda\ll x\leq \xid}{\sim}
\soma \ea \roa^{{\rm cl}(\eps)}(x) +
 \hbar \, \B \frac{\kd^2}{4 \pi} e^{- \kd x}, 
\nonumber
\end{equation}
where, by virtue of \eqref{valuerhocl} and \eqref{rappelPhicl}, 
\begin{equation}
\soma \ea \roa^{{\rm cl}(\eps)}(x)=
- \frac{1}{2} 
    \kd \beta\left( \somg \eg^3 \rogb\right) \left[ \, \Lb (\kd x) - \Mb (\kd x)
    \right]. 
\end{equation} 
 In the case of a charge-symmetric two-component plasma 
$\soma \ea \roa (x)$ is purely quantum at distances $x\gg \lambda$.

Eventually, the large-distance behavior of $\soma \ea \roa (x)$ 
is merely ruled by the electrostatic potential drop $\Phi(x)$,  which includes both 
quantum and classical effects,
\begin{equation}
\soma \ea \roa (x) \underset{x \gg \xid}{\sim} - \frac{\kd^2}{4 \pi}
\left[\Pas^{{\rm cl}(\eps)} -\hbar\B\right] \, e^{- \kd x},
\end{equation}
where $\Pas^{{\rm cl}(\eps)}$ is given in \eqref{valuePhias}.
 As it is the case for  the 
electrostatic potential drop and for particle densities, the charge density
includes quantum effects, linear in $\hbar$, which exist
far away from the wall over a few Debye screening lengths. 

\subsection{Global charge}
\label{section64}

The wall that we consider does not carry any external surface charge and  the global surface electroneutrality \eqref{neutrsurf} of the Coulomb fluid 
 is fulfilled, as checked in the present section. 
At leading order $\cO\left( \eps,\kzl\right)$ the global surface charge $\sigma$ can be decomposed into leading  classical and  quantum contributions:
$\sigma=\sigma^{{\rm cl}(\eps)}+\sigma^{{\rm qu}(\kzl)}$, where
\begin{equation}
\sigma^{{\rm cl}(\eps)}\equiv\int_0^{+\infty} dx\soma \ea
\roa^{{\rm cl}(\eps)}(x),
\end{equation}
and $\sigma^{{\rm qu}(\kzl)}$ is the term of order $\kzl$ in 
\begin{equation}
\label{defsigmaqu}
\int_0^{+\infty} dx \soma \ea
\left[\roa(x)-\roa^{{\rm cl}(\eps)}(x)\right].
\end{equation}
We recall that, as mentioned after \eqref{valuerhocl}, the hard-core that must be introduced in order to prevent the classical collapse between opposite charges does not appear at order $\eps$ in $\roa^{\rm cl}(x)$.
 
Since the classical  densities already obey 
\eqref{neutrsurf}, which is  enforced by macroscopic electrostatics, 
$\sigma^{{\rm cl}(\eps)}$ and $\sigma^{{\rm qu}(\kzl) }$ must vanish separately.
As checked in Ref.\cite{jn&fr01I},
the classical contribution $\sigma^{\rm cl (\eps)}$ of order $\eps$ arising from \linebreak
$\soma \ea\roa^{{\rm cl}(\eps)}(x)$ does vanish.

The
quantum term 
$\sigma^{{\rm qu} (\kzl)}$ of order $\kzl$ arises only from
the two quantum terms 
$- \roab \exp\left[-2x^2/\laa^2\right]$ and 
$- \roab\beta\ea\Phi^{{\rm qu}(\kzl)}(x)$ in $\roa(x)-\roa^{{\rm cl}(\eps)}(x)$.
Indeed, according to \eqref{relrhoqurhocl},
\begin{eqnarray}
\label{decomprho}
\roa(x)-\roa^{{\rm cl}(\eps)}(x)=
&&-\roab e^{-2x^2/\laa^2}-\roab \beta \ea \Phi^{{\rm qu}(\kzl)}(x)\\
&&-\left\{
\left[\roa^{{\rm cl}(\eps)}(x)-\roab\right]
-\roab\beta \ea \Phi^{{\rm qu}(\kzl)} (x)
\right\}e^{-2x^2/\laa^2}.\nonumber
\end{eqnarray}
The term in  curly brackets is a function of order  $\roab\cO( \eps, \kzl)$ which decays exponentially fast over the scale $\kz^{-1}$; by virtue of  
\eqref{prop2bis}, after multiplication by 
$\exp\left[-2x^2/\laa^2\right]$ this term gives a contribution of relative order $\cO\left( \eps\,.\,\kzl,(\kzl)^2 \right)$ to $\sigma^{{\rm qu}(\kzl)}$ defined in \eqref{defsigmaqu}. Therefore \linebreak
$\sigma^{{\rm qu} (\kzl)}=\sigma^{{\rm qu} (\kzl)}_{<}
+\sigma^{{\rm qu} (\kzl)}_{>}$  with
\begin{equation}
\label{sigmaW}
\sigma^{{\rm qu} (\kzl)}_{<}\equiv -
\int_0^{+\infty} dx \, \soma \ea \roab  e^{- \dxdslad}
=-\hbar\, \frac{1}{2}\sqrt{\frac{\pi \beta}{2}}
\soma \frac{\ea}{\sqrt{\ma}}\roab 
\end{equation}
and
\begin{equation}
\label{sigmaphiqu}
\sigma^{{\rm qu} (\kzl)}_{>}\equiv -
\int_0^{+\infty} dx \, \soma \ead \roab  \beta \Phiquu (x),
\end{equation}
where $ \Phiquu (x)$ is
given in \eqref{valuePhiqubis}.
 
 The $\hbar$-contribution
$\sigma^{{\rm qu} (\kzl)}_{<}$  to the global charge $\sigma$  
is canceled by the contribution $\sigma^{{\rm qu} (\kzl)}_{>}$. Therefore, the quantum term $\beta \ea \Phiquu (x)$ in
density  profiles, which  varies over the classical Debye screening 
length, can be seen as being enforced by the interplay between 
the global surfacic electroneutrality condition and the  
fact that, when  wave-functions vanish over different length scales, a charge-density profile appears
 in the very vicinity of the wall even in the zero-coupling limit (see \eqref{neutrsurfid}). 

 The previous calculation can also be interpreted as follows. We notice that $\sigma^{{\rm qu} (\kzl)}_{<}$ and 
$\sigma^{{\rm qu} (\kzl)}_{>}$ can be viewed as the leading $\hbar$-terms in the contributions to $\sigma^{\rm qu}$ from the regions $x<l$ and $x>l$ respectively, with $\lambda\ll l\ll \xid$. Indeed,
\begin{equation}
\label{sigmaloc}
\sigma^{{\rm qu} (\kzl)}_{<}=
\mathop{\lim}_{(\lambda/l)\rightarrow 0}\,
\mathop{\lim}_{(l/\xid)\rightarrow 0}
\left(\int_0^{l} dx \, \soma \ea
 \left[\roa(x)-\roa^{{\rm cl}(\eps)}(x)\right]\right)^{(\kzl)},
\end{equation}
whereas 
\begin{equation}
\label{sigmaetend}
\sigma^{{\rm qu} (\kzl)}_{>}=
\mathop{\lim}_{(\lambda/l)\rightarrow 0}\,
\mathop{\lim}_{(l/\xid)\rightarrow 0}
\left(\int_l^{+
\infty} dx \, \soma \ea 
 \left[\roa(x)-\roa^{{\rm cl}(\eps)}(x)\right]\right)^{(\kzl)},
\end{equation}
where $\roa(x)-\roa^{{\rm cl}(\eps)}(x)$ is given in 
\eqref{decomprho}.
$l$ can be identified with the mean interparticle distance $a$, by virtue of \eqref{encadra}.
Therefore, since $\lambda\ll a$,
$\sigma^{{\rm qu} (\kzl)}_{<}$ can be seen as a surface charge located in the plane $x=0$, whereas
$\sigma^{{\rm qu} (\kzl)}_{>}$ is spread in the fluid over the scale $\xid$. On the other hand, 
according to \eqref{Phiquvalue} and \eqref{sigmaW}, $ \Phiquu (x)$ may be written as
\begin{equation}
 \Phiquu (x)=\int d\vecr' 
\sigma^{{\rm qu} (\kzl)}_{<} \delta(x')\, \phi^{(0)}(\vecr,\vecr').
\end{equation} 
The interpretation of the latter equation is that $\Phiquu (x)$ is the classically-screened electrostatic potential created by the part of the fluid charge-density profile which is concentrated near the wall.


In the
case of  an intrinsic semiconductor  near a junction,
 the system of electrons and
positive holes in the conduction band can be considered as a
two-component Coulomb fluid of charges $-q_e$ and $+q_e$ embedded in a medium of relative
dielectric constant $\epsilon_{\rm m}$.  $q_e$ is the absolute value of the electron charge and  energy terms  involve
$e\equiv q_e/\sqrt{\epsilon_{\rm m}}$ (see the comment after \eqref{defv}). Since the semiconductor is intrinsic,
the densities $\rho_-$ and
$\rho_+$ are equal to each other. They are determined from the energy gap $E_{G}$ and from the effective masses 
$m^{\rm eff}_-$ and $m^{\rm eff}_+$ by  \cite{AshMermin} 
\begin{equation}
\rho=\rho_{\pm}(\beta) = \frac{1}{4}
\left( \frac{2}{\pi \beta \hbar^2}\right)^{3/2} \,
\left( m^{\rm eff}_- \, m^{\rm eff}_+ \right)^{3/4} \,
e^{- \beta E_G /2}.
\end{equation}
Since the system is charge-symmetric, there is no classical contribution to the potential drop $\Phi(x)$ and
\eqref{qdensfnch} becomes
\begin{equation}
\label{semicond}
\rho_{\pm}(x)=\rho\left(1-e^{-2x^2/\lambda_{\pm}^2}\right)
\left[1-\eps \Lb (\kd x) \mp \beta 
\frac{q_e}{\sqrt{\epsilon_{\rm m}}}\Phi^{\rm qu}(0) e^{-\kd x}\right],
\end{equation} 
where $\eps\equiv\beta q_e^2/(2\epsilon_{\rm m})$ and, according to 
\eqref{valuePhiqubis} and \eqref{valueB},
\begin{equation}
\label{semibis}
\beta 
\frac{q_e}{\sqrt{\epsilon_{\rm m}}}\Phi^{\rm qu}(0)= 
\frac{1}{4}\sqrt{\frac{\pi}{2}}
\kd\lambda_-\left[1-
\sqrt{\frac{m^{\rm eff}_-}{m^{\rm eff}_+}}\right].
\end{equation} 
(We notice that on principle the expression \eqref{semicond} is valid only when the wall has the same dielectric constant as the medium where charge carriers move.)

In the case of GaSb, $E_G=0.67 \, eV$ at $273 \, K$,
$m^{\rm eff}_-=0.047 \, m_{e}$, $m^{\rm eff}_+=0.5 \, m_{e}$ (where $m_e$ is the electron mass), and $\epsilon_{\rm m}=15$. This system is 
  in the regime \eqref{regimed} for which explicit analytical expressions are
calculated in the present paper: 
 $\kd \lambda_-= 2.5\, 10^{-3} $,
$\eps=6\, 10^{-4}$, 
$(\lambda_-/a)^3=2\, 10^{-6} $ (with $(4/3)\pi a_-^3\rho_-=1$). (The length scales are 
$\lambda_+\sim 2.5\, {\rm nm}$, $\lambda_-\sim 8.3\,{\rm nm}$, 
$a\sim 640\, {\rm nm}$ and $\xid\sim 3300\, {\rm nm}$.)
The vanishing of particle densities  occurs on two
 different scales $\lambda_+$ and $\lambda_-\sim 3.3 \lambda_+$. According to \eqref{sigmaW}, where $\ea=\pm q_e$, the resultant surface charge  located on the wall (over the width $\lambda_-$) is $\sigma^{{\rm qu} (\kzl)}_{<}=5\, 10^{-14} C {\rm cm}^{-2}=3\, 10^{5} q_e\, {\rm cm}^{-2}$.  The bulk density  of charge carriers is $\rho_-+\rho_+\sim 2\, 10^{12}\, {\rm cm}^{-3}$ and the  charge density at distances  $x>a$ is $\rho_c \exp[-\kd x]$ with, according to \eqref{semicond} and \eqref{semibis}, $\rho_c =10^9\,q_e {\rm cm}^{-3}$. Thus  $\rho_c\lambda_-\sim 1\,q_e\, {\rm cm}^{-2}$ is indeed negligible compared with
$\sigma^{{\rm qu} (\kzl)}_{<}$.
The  potential drop $\Phi^{\rm qu}(0)=1.3\,  10^{-5} eV$ remains negligible with respect to the energy gap.


\section{Comment}
\label{section7}

In this section, we comment on   the case where 
the wall is  made of a dielectric material, characterized by a relative 
dielectric constant $\ew $ with respect to the vacuum, when $\ew$ is different from the relative dielectric constant $\epsilon_{\rm m}$ of the medium where charges move. 
Then the Coulomb interaction reads
\begin{equation}
\label{defvw}
v_{\scriptscriptstyle W}(\vecr,\vecr')
=\frac{1}{|\vecr-\vecr'|}-\De\frac{1}{|\vecr-\vecr'^{\star}|}
\end{equation}
with $\De = (\ew -\epsilon_{\rm m})/(\ew +\epsilon_{\rm m})$.
The response of the wall induced by the presence of a particle 
with charge $\ea$ (which includes a factor 
$1/\sqrt{\epsilon_{\rm m}}$ in interaction terms) is equivalent to the presence of an image charge at position $\vecr^{\star}$, symmetric
of the real-particle position $\vecr$ with respect to the wall, and which carries a
charge $-\De \, \ea$.
The Hamiltonian also  involves the self-energy $- \De \, \ead / 4 x$, 
due to the interaction of a particle with its own image charge.
The corresponding loop self-energy can be incorporated in the loop fugacity $z(\Lr)$, which now reads
\begin{equation}
\label{zselfloop}
z(\Lr)=\za \theta(x)
\exp\left[
\De \frac{\beta \ea^2}{4} \int_0^1 ds \frac{1}{x+\laa\xi_x(s)}\right].
\end{equation}

In order to exhibit the screening of the self-energy, which is not integrable at large distances $x$ from the wall,  we 
have performed a resummation in two steps, which generalizes  the 
method devised for classical systems in Ref.\cite{jn&fr01II}.
The choice of the same nine resummed bonds as 
those in Section \ref{section32} combined with the two-step resummation leads to resummed  weights which  are integrable, because they involve only screened loop-fugacities.

Indeed, in the one-step resummation of Section \ref{section32} the weights corresponding to the nine resummed bonds are 
$\zs(\Lr)$ and
$\zs(\Lr) - z(\Lr)$, instead of the weights 
$z(\Lr)$ and $\zs(\Lr) - z(\Lr)$ that arise when there are only five resummed bonds without any ``{double}'' bond, such as $\left[\Fcc\right]^2/2$, as it is done in Ref.\cite{Cornu96I}.
At the end of the two-step resummation process, the construction rules for resummed diagrams are the same ones as in the  one-step resummation of Section 3.2, with the only difference that weights 
$\zs(\Lr)$ and
$\zs(\Lr) - z(\Lr)$ are replaced by weights $z^{{\rm sc}[2]}(\Lr)$ and $z^{{\rm sc}[2]}(\Lr)-z^{{\rm sc}[1]}(\Lr)$. These weights are integrable  at large distances $x$ from the wall, because both $z^{{\rm sc}[2]}(\Lr)$ and $z^{{\rm sc}[1]}(\Lr)$  result from resummations of  Coulomb ring subdiagrams.

More precisely, the expressions of the resummed loop fugacities $z^{{\rm sc}[i]}(\Lr)$
(with $i=1,2$) are  given by \eqref{defzs} and \eqref{defVs} where the value of $z(\Lr)$ is that given in \eqref{zselfloop} and $\phi-v$ in $\calVs(\Lr) $ is replaced by  
$\phi^{[i]}_{\scriptscriptstyle W}-v_{\scriptscriptstyle W}$;
the $\phi^{[i]}_{\scriptscriptstyle W}$'s  have the same boundary conditions as $v_{\scriptscriptstyle W}$ written in \eqref{defvw} and they obey the inhomogeneous Debye equation \eqref{eqgene} where
${\overline{\kappa^{[1]}}}^2 (x)  \equiv \, 4 \pi \beta \sum_{\alpha}\ea^2 \int \Dvxi \, z(\Lr)$ and
 ${\overline{\kappa^{[2]}}}^2 (x)  \equiv \, 4 \pi \beta \sum_{\alpha}\ea^2 \int \Dvxi \, z^{{\rm sc}[1]}(\Lr)$. (Resummed bonds are defined with $\phi^{[2]}_{\scriptscriptstyle W}$ in place of $\phi$.)

As a consequence of  quantum dynamics, screened loop self-energies
(and subsequently particle densities) are found to approach their bulk values only with an integrable  $1/x^3$ tail, whereas
the particle self-energy due to the electrostatic response of the wall is
 exponentially screened  in classical systems  
\cite{Onsa&Sama34,jn&fr01I}. Even for
 bulk properties 
\cite{Angel&Phil89,Cornu96II}, screening in quantum systems is
less efficient than in classical fluids.

More precisely, the screened loop self-energy in
$z^{{\rm sc}[1]}(\Lr)$ or $z^{{\rm sc}[2]}(\Lr)$ is the sum of two contributions. 
The  exponentially-decaying part   has the same decay 
at large distances from the wall as the  screened self-energy of a classical charge \cite{jn&fr01II}.
The algebraic $1/x^3$ tail   arises from the other part, which reads
\begin{equation}
\label{screenedself}
\De\frac{\ea^2 }{2} \int_0^1ds\int_0^1 ds' \Bigl(1-\delta(s-s')\Bigr)
\frac{1}
{\vert\vecr+\laa\vxi(s) -\vecr^{\star}-\laa\vxi^{\star}(s')\vert}
\end{equation} 
(See the analogous term \eqref{defW} in the screened pair interaction). 
In the  low-degeneracy and weak-coupling regime
the particle density $\roa(x)$ does not seem to have a simple explicit value, because of the self-energy contributions  arising from the dielectric response of the wall.

\appendix

\section{: Resummation of  large-distance Coulomb divergences}

\label{appresum}

In this appendix, we display the resummation of Coulomb divergences
in the Mayer representation \eqref{expmayrq} for the loop-fugacity expansion of the loop density. 
We use the same decomposition of $f$-bonds as the one  performed in
Appendix B of Ref. \cite{Cornu96I}. (The {\it a priori} arbitrary decomposition
is chosen according to the properties to be studied after resummation.) 
The $f$-bond is  written as the sum of ten auxiliary bonds $\tilde{f}$,
\begin{multline}
  \label{decompfq}
  f(\Lr, \Lr') = \left\{ \fcc + \fmc + \fcm + \fmm  
\phantom{\frac{1}{2} \left[ \fcc \right]^2}  \right.\\
\left. + \frac{1}{2}
   \left[ \fcc \right]^2 + \fcc \fmc + \fcc \fcm + \frac{1}{2} \left[
   \fmc \right]^2 + \frac{1}{2} \left[ \fcm \right]^2 + \ftt \right\} 
 (\Lr, \Lr'). 
\end{multline}
Bonds $\fcc$,  $\fmc$, $\fcm$ and  $\fmm$ are defined from the multipolar
decomposition \eqref{decompvwq} and are introduced in order to handle
classical exponential screening.  These bonds read
\begin{equation}
  \label{sumfaux}
  f^{{\rm a \, b}} (\Lr, \Lr') =  - \beta e_{\alpha} e_{\alpha'} \, 
 {\mathcal{V}}^{{\rm a \, b}} (\Lr, \Lr'),
\end{equation}
where superscripts $a$ 
and $b$ stand either for ${\rm c}$ or ${\rm m}$, and where 
we have set
${\mathcal V}^{{\rm c \, c}} (\Lr, \Lr') \equiv v(\vecr - \vecr')$,. 
$\ftt$ is defined by \eqref{decompfq}. 
Double $f$-bonds (such as $\left[\fcc \right]^2/2$, $\fcc \fcm$, ...) are 
not involved in the series representation \eqref{expmayrq} where 
two points can be linked by at most one $f$-bond. 
We choose to make
them appear  in the decomposition \eqref{decompfq} in order to 
obtain  a finite sum for the contribution from all possible  ring subdiagrams  
which are defined  hereafter.
When the decomposition \eqref{decompfq} is introduced in the 
Mayer representation \eqref{expmayrq} of the loop density, diagrams 
$\cg$ are replaced by diagrams  $\tilde{\cg}$ built with the ten $\tilde{f}$-bonds 
defined in \eqref{sumfaux} and the same topological rules as diagrams $\cg$.

The resummation procedure  relies on the integration over the 
intermediate points (called ``{Coulomb points}'') of all possible Coulomb-chain
or Coulomb-ring subdiagrams.
A  Coulomb-chain subdiagram between two points $\Lr$ and $\Lr'$ is equal to  
\begin{multline}
\label{defchainbis}
\int d\Lr_1\cdots d\Lr_N \, 
f^{{\rm a \, c}} (\Lr, \Lr_1) \, z(\Lr_1) \, \fcc (\Lr_1, \Lr_2) 
\, z(\Lr_2) \, \fcc (\Lr_2, \Lr_3) \\
\times\cdots z(\Lr_N) \, f^{{\rm c \, b}} (\Lr_N, \Lr')
\end{multline}
 with an arbitrary number $N\geq 1$ of internal points.
A Coulomb-ring subdiagram attached to a point $\Lr$ is a
closed  Coulomb-chain subdiagram, i.e. with $\Lr = \Lr'$ (see Fig.\ref{Ir}). It is equal either to 
$\int d\Lr_1 \, z(\Lr_1) \left[\fcc(\Lr,\Lr_1)\right]^2 / 2$,  \linebreak
$\int d\Lr_1 \, z(\Lr_1)\left[\fmc(\Lr,\Lr_1)\right]^2 / 2$ ,   
$\int d\Lr_1 \, z(\Lr_1) \fcc(\Lr,\Lr_1)\fmc(\Lr,\Lr_1)$ or 
\begin{multline}
\frac{1}{S}\int d\Lr_1\cdots d\Lr_N \, 
f^{{\rm a \, c}} (\Lr, \Lr_1) 
\, z(\Lr_1) \, \fcc (\Lr_1, \Lr_2) \,  
z(\Lr_2) \, \fcc (\Lr_2, \Lr_3)\\
\times\cdots  z(\Lr_N) \, f^{{\rm c \, b}} (\Lr_N, \Lr)
\end{multline}
where $S$ is the symmetry factor of the Coulomb-ring subdiagram ($S=2$ if $a=b$ and $S=1$
otherwise). Coulomb-ring subdiagrams do exist in diagrams $\tilde{\cg}$,
because the latter ones contain articulation points.

Diagrams $\tilde{\cg}$ can be collected into classes of a partition where all diagrams 
$\tilde{\cg}$ inside the same class lead to  the same prototype diagram
$\cpps$ after erasing the intermediate points of all Coulomb-chain or 
Coulomb-ring subdiagrams. When resummed bonds are defined, some ``{excluded-composition}'' rules 
ensure a one-to-one correspondence between each class in the partition of diagrams $\tilde{\cg}$ 
and each diagram $\cpps$. In the present paper, we choose to build prototype diagrams 
$\cpps$ made with bonds $\Fcc$, $\Fcm$, $\Fmc$, $(1/2)\left[\Fcc\right]^2$, $(1/2)\left[\Fcm\right]^2$, 
$(1/2)\left[\Fmc\right]^2$,  $\Fcc\Fcm$, $\Fmc\Fcc$ and $\Frt$ (see \eqref{defF}). 
This choice leads to resummed diagrams where the new weight of every point is 
convenient for dealing with the case of a wall with a dielectric response, as explained in Section \ref{section7}.

 The bond $\Fcc(\Lr,\Lr')$ between two loops $\Lr$ and 
$\Lr'$ of a prototype diagram $\cpps$ is defined as the sum of 
the bond $\fcc(\Lr,\Lr')$ and of  all Coulomb chains involving only $\fcc$-bonds 
between $\Lr$ and  $\Lr'$ (see Fig. \ref{figphi}).
The bond $\Fmc(\Lr,\Lr')$ is the sum of the bond $\fmc(\Lr,\Lr')$ and 
of all Coulomb chains \eqref{defchainbis} with $a={\rm m}$ and $b={\rm c}$ 
(see Fig.\ref{figFmc}). 
Bonds $\left[\Fcc \right]^2/2$, $\left[\Fcm \right]^2/2$, 
$\left[\Fmc \right]^2/2$, $\Fcc \Fcm$ and  $\Fcc \Fmc$ between two 
points $\Lr$ and $\Lr'$ originate from subdiagrams with either one bond 
$(1/2)\left[\fcc\right]^2$, $(1/2)\left[\fcm\right]^2$, \linebreak
$(1/2)\left[\fmc\right]^2$,  $\fcc\fcm$, or $\fcc \fmc$ respectively,  or from subdiagrams with two Coulomb chains 
between $\Lr$ and $\Lr'$ (see Fig.\ref{figFccd}). 
In Ref.\cite{Cornu96I}, the prototype diagrams were chosen to be built with only five resummed bonds called
$\Fcc$, $\Fcm$, $\Fmc$ and $F_{{\rm\scriptscriptstyle R}_z}$ and 
$F_{\rm{\scriptscriptstyle R T}_z}$.  Since the resummation procedure relies 
on the integration over the same intermediate points in the same  diagrams $\tilde{\cg}$ in both  cases, the
expressions of $\Fcc$ and $\Fmc$ are the same (see Eqs. \eqref{defFcc} and \eqref{defFmc}), 
while  $\Frt$ is equal to
\begin{equation}
\Frt=F_{{\rm\scriptscriptstyle R}_z}
 - \frac{1}{2} \left[\Fcc \right]^2 - \frac{1}{2} \left[\Fcm \right]^2 -
 \frac{1}{2} \left[\Fmc \right]^2 - \Fcc. \Fcm - \Fcc. \Fmc.
\end{equation}
>From the expression of $F_{{\rm\scriptscriptstyle R}_z}$ derived in 
Ref. \cite{Cornu96I}, we get the value \eqref{defFr}.

The ``{excluded-composition}'' rule is different from those of 
Ref.\cite{Cornu96I}, because  resummed bonds are different in the 
present case. The rule arises from the following arguments. When 
$\Lr$ is the intermediate point of the chain 
$F^{\rm ac}(\Lr_i,\Lr) \, F^{\rm cb}(\Lr,\Lr_j)$ with $i\not=j$ and
is not involved in any other bond, or when $\Lr$ appears 
only in one bond
$(1/2)\left[F^{\rm ac}(\Lr_i,\Lr)\right]^2$ (with $a={\rm c}$ or ${\rm m}$) 
or one bond $\Fcc(\Lr_i,\Lr)\Fmc(\Lr_i,\Lr)$, the reason why  $\Lr$ has  not disappeared
in the resummation process can only be that $\Lr$ carried at least one Coulomb ring in 
every diagram $\tilde{\cg}$ in the class which leads to the diagram
$\cpps$ after erasing  Coulomb points. 
Therefore, after  integration over Coulomb points, the weight $z(\Lr)$ attached to point $\Lr$ in diagrams $\tilde{\cg}$ is 
multiplied in diagrams $\cpps$ by the sum of all products of Coulomb rings attached to $\Lr$. 
As argued in  Ref.\cite{Cornu96I}, the sum is equal to $\exp I_{\rm r}-1$, 
where $I_{\rm r}$ is the sum of all possible Coulomb rings (see Fig. \ref{Ir}).
The corresponding weight  is equal to 
$\zs(\Lr)- z(\Lr) $ where $\zs(\Lr) \equiv z(\Lr) \exp {\Ir}$.
$\Ir$ is linked 
to the screened potential 
$\phi$ by $\Ir = - \beta \ead \calVs$ where $\calVs$ is 
defined in \eqref{defVs}.
$\zs(\Lr)$ exhibits the stabilizing effect of screening clouds
 \cite{jn&fr01II}. 
 When $\Lr$ is not involved only in a product 
$F^{\rm ac}(\Lr_i,\Lr)F^{\rm cb}(\Lr,\Lr_j)$ (where $\Lr_i$ may coincides with $\Lr_j$), $\Lr$ may carry no Coulomb ring in the 
 diagrams $\tilde{\cg}$  of the corresponding class, and its weight in diagrams
$\cpps$ is equal to $\zs(\Lr)$.



\section{: Scaling analysis}

\label{appscal}

In this appendix, we perform the scaling analysis of  diagrams 
with one internal point in the resummed Mayer expansion \eqref{expmayrqf}.

\subsection{$\Fmc$- and $\Fcm$-bonds}

 A simple scaling analysis of the contribution from diagrams with either one  bond $\Fcm(\Lr, \Lr')$ or one bond $\Fmc(\Lr, \Lr')$ is obtained by a Taylor expansion of their expressions  (see \eqref{defFmc}) around their classical values.

In the case of diagram with  one bond $\Fmc$  the 
Taylor expansion in the variable $\lambda_{\alpha} \vxi (s) $ can be factorized: the leading order is given by 
\begin{equation}
\int_0^1 ds 
 \lambda_{\alpha} \vxi (s) . \nabla_{\vecr} \left( 
 \int d\Lr' \Fcc(\Lr, \Lr') \, \, \zs ( \Lr') \right).
\end{equation} 
 According to the discussion in Section \ref{section53}, at leading order the parenthesis 
proves to be a function of $\vecr$ which varies over the scale $\kz^{-1}$.
By using the fact that 
the action of the gradient 
$\lambda_{\alpha}  \nabla_{\vecr}$ on a function of $\kz \vecr$ multiplies its order by $ \kz \lambda_{\alpha}$, 
we get from \eqref{orderdiagFcc} that
\begin{equation}
  \label{orderFmc}
  \int d\Lr' \Fmc(\Lr, \Lr') \, \, \zs ( \Lr') = 
\cO\Bigl(\eps. \kzl, \left(\kzl\right)^2\Bigr).
\end{equation}

The diagram with one bond $\Fcm(\Lr, \Lr')$ involves integrated moments of $\vxi'$, which depend on $\lag$, and $\somg \eg \zg$ can no longer be factorized at leading order. Moreover,  the first moment of the $x$-component of $\xi$ does not vanish, and the  order of 
 $\int d\Lr' \Fcm(\Lr, \Lr') \, \, \zs ( \Lr')$ is given by 
\begin{equation}
  \label{eqcsi}
  \somg  \zg  \int d\vecr' \, 
\left(  \int_0^1 ds \,  \int \Dvxigp \, \xi'_{x'}(s) \right)
 \lag \frac{\partial \Fcc (\vecr, \vecr')}{\partial x'}
\end{equation}
At leading order $\lag \partial\Fcc (\vecr, \vecr')/\partial x'$ is a function of $\kz \vecr$ and $\kz \vecr'$, while, 
as can be checked in \eqref{intmesqx}, the mean extension of 
a Brownian path, $\int_0^1 ds \,  \int \Dvxigp \, \xi'_{x'}(s)$, 
vanishes gaussianly fast when $x'/\lag$ goes to infinity.
The latter function makes the next $x'$-integration  convergent over 
the scale $\lambda$ (and not $\kz^{-1}$). Therefore, as in \eqref{prop2bis}, the order of the integral \eqref{eqcsi} is equal to 
$\kzl$ times the order of 
$ \zg \int d\vecr' \, 
\lag \partial\Fcc (\vecr, \vecr')/\partial x'$.
Since at leading order $\Fcc (\vecr, \vecr') $ is a function of $\kz \vecr$ and $\kz \vecr'$, 
the latter integral
 is of order $\kzl$ times the order of 
$ \zg \int d\vecr' \, \Fcc (\Lr, \Lr')$, which is equal to
$\cO(1)$
by virtue of \eqref{prop0}. Eventually
\begin{equation}
  \label{orderFcm}
  \int d\Lr' \Fcm(\Lr, \Lr') \, \, \zs ( \Lr') = 
\cO\left(\left(\kzl\right)^2\right).
\end{equation}

\subsection{Multiple bonds}

In the case of multiple bonds, the dependence upon the species of the loop $\Lr'$ is not reduced to $\eg\zg$. Therefore,  
 contrary to what happens in the case of $\Fcc$,  the neutrality constraint  on fugacities \eqref{zneutr} no longer increases the actual order with respect to that given by pure scaling analysis. For instance, since $\left[\Fcc\right]^2$ is proportional to $\eps^2\left[\phitzero(\vecrt,\vecrt')\right]^2$, while $\Fcc$ is proportional to 
$\eps\phitzero(\vecrt,\vecrt')$,
\eqref{prop0} implies that 
\begin{equation}
\label{prop3}
\int d\Lr' \, \frac{1}{2}\left[\Fcc\right]^2 (\Lr, \Lr')
\zs(\Lr') =\cO(\eps).
\end{equation}
However, owing to the construction rule \eqref{poidsres}
about fugacities, the internal point in the diagram made with one 
bond $\left[\Fcc\right]^2/2$ has a weight equal to $\left[\zs(\Lr') - z(\Lr')\right]$. 
By virtue of \eqref{dimanche}, the latter weight is of order 
$\cO(z \eps)$ and
 \eqref{prop3} leads to
\begin{equation}
  \label{trueFcccarre}
  \int d\Lr' \frac{1}{2} \left[\Fcc \right]^2 (\Lr, \Lr')
\, \, \left[ \zs (\Lr') - z(\Lr')\right]  = \cO(\eps^2 ).
\end{equation}

Similarly diagrams $\left[\Fcc \Fmc\right] (\Lr, \Lr') $ and $\frac{1}{2} \left[\Fmc\right]^2 (\Lr, \Lr') $ also have an internal point which carries the weight 
$\zs(\Lr')-z(\Lr')$. Only the moments of the Brownian path $\vxi$, which is associated with the nonintegrated position $x$,  are involved and,  according to the argument leading to \eqref{orderFmc}  and by virtue of \eqref{trueFcccarre}, the  contributions from these diagrams are 
\begin{equation}
  \label{}
  \int d\Lr'  
\left[\Fcc \Fmc \right]\Lr, \Lr') \, \,
\left[ \zs (\Lr') - z(\Lr')\right] =\cO\left(\eps^2 \,.\,\kzl\right)
\end{equation}
and
\begin{equation}
  \label{}
  \int d\Lr' \frac{1}{2} \left[\Fmc \right]^2 (\Lr, \Lr')
\, \, \left[ \zs (\Lr') - z(\Lr')\right]  = 
\cO\left(\eps^2 \,.\,\left(\kzl\right)^2\right).
\end{equation}

For the bond $\left[\Fcc \Fcm \right]$, by virtue of the  mechanism involved in \eqref{orderFcm}, the integration over $x'$  of the first moment of the Brownian path $\vxi'$ leads to an extra factor $\left(\kzl\right)^2$ with respect to the order \eqref{prop3} of 
the contribution from $\left[\Fcc \right]^2/2$, namely
\begin{equation}
  \label{}
  \int d\Lr' \left[\Fcc \Fcm \right]
(\Lr, \Lr') \, \, \zs ( \Lr') = \cO\left(\eps\,.\,\left (\kzl\right)^2 \right).
\end{equation}

At leading order the bond $\left[\Fcm \right]^2/2 $ involves the second moment of the Brownian path $\vxi'$. This moment tends to a non-zero value when $x$ goes to infinity, so that the integration over $x'$ still converges over the scale $\kz^{-1}$. Therefore, according to the argument leading to \eqref{orderFmc},  the order of the contribution from $\left[\Fcm \right]^2/2 $ has  an extra factor of order $(\kzl)^2$ with respect to  \eqref{prop3}, 
\begin{equation}
  \label{}
  \int d\Lr' \frac{1}{2} \left[\Fcm \right]^2 
(\Lr, \Lr') \, \, \zs ( \Lr') = \cO\left(\eps\,.\,(\kzl)^2 \right).
\end{equation}

\subsection{$\Frt$-bond}

Let us consider the diagram made of a single bond  $\Frt$. The integration over the Brownian 
path $\vxi'$ makes the integral over $\vecr'$ convergent
at small distances $|\vecr - \vecr'|$. The order of the contribution from this diagram can be inferred from the known results  about the bond $\Frt^{\star}$ obtained by adding to $\Frt$ the double bonds different from  $\left[\Fcc\right]^2/2$, which do not contribute to  the integrability of 
$\Frt$ at large  distances $|\vecr - \vecr'|$ in the limit where  $\eps$ vanishes.  $\Frt^{\star}$ is defined by 
\begin{equation}
\Frt=\Frt^{\star}
 - \frac{1}{2} \left[\Fcm \right]^2 -
 \frac{1}{2} \left[\Fmc \right]^2 - \Fcc. \Fcm - \Fcc. \Fmc.  
\end{equation} 
where $\Frt$ is given in \eqref{defFr}.

The order of the contribution of 
$\int d\vecr' \int \Dvxi \int \Dvxip \, \Frt^{\star} (\Lr, \Lr')$ 
has been studied in the bulk situation in Ref.\cite{Cornu98II} (and a similar calculation also appears in Ref.\cite{ACP95}). Let us introduce the thermal de Broglie wavelength $\lambda_{\alpha\gamma}$
associated with
 the reduced mass 
$m_{\alpha} m_{\gamma}/(m_{\alpha} +m_{\gamma})$. In the bulk, the considered integral  proves to be the sum of a term of order $\lambda_{\alpha\gamma}^3$ times an analytic
 function $Q(\xi_{\alpha\gamma})$ of the   parameter 
 $\xi_{\alpha\gamma}\equiv -\beta \ea\eg/\lambda_{\alpha\gamma}$  \cite{Ebeling69}, 
 plus a term of order $(\beta \ea\eg) \lambda^2$ (with two contributions where $\lambda=\lambda_{\alpha}$ or 
$\lambda=\lambda_{\gamma}$). The latter ``{diffraction}'' term arises from the second moments of the Brownian paths. There is no term proportional to  $(\beta \ea\eg)^2 \lambda$, because the first moment of a Brownian path vanishes in the bulk, by virtue of the spherical symmetry.

In the vicinity of the wall, the precise calculation is more delicate than in the bulk. However, we may expect to obtain the same  orders $\lambda_{\alpha\gamma}^3$ and  
$(\beta \ea\eg) \lambda^2 $ as in the bulk, plus a term of order 
$(\beta \ea\eg)^2 \lambda$ allowed by the anisotropy introduced by the presence of the wall.
Since the fugacity $\zsc (\Lr')$ is
of order $\cO(\rho) = \cO(a^{-3})$, the  contribution from the diagram made of a single $\Frt$-bond 
can be viewed as the sum of four terms with respective orders
\begin{equation}
\left(\frac{\lambda}{a}\right)^
3Q_{\scriptscriptstyle W}\left(-
\frac{\beta \ea \eg}{\lambda_{\alpha\gamma}}\right), 
\end{equation} 
where $Q_{\scriptscriptstyle W}$ is defined similarly to $Q$ mentioned previously \cite{Ebeling69} with  the only difference that space integrals are restricted to $x>0$ and $x'>0$,
\begin{equation}
\left(\frac{\beta \ea \eg}{a} \right)
\left(\frac{\lambda}{a}\right)^2  =\cO \left( (\kzl)^2 \right),
\end{equation}
\begin{equation}
\left(\frac{\beta \ea \eg}{a} \right)^2
\left(\frac{\lambda}{a}\right) 
=\cO \left( \eps\,.\,\kzl \right),
\end{equation}
and
 \begin{equation}
\left(\frac{\beta \ea \eg}{a} \right)^3
\ln (\kzl) 
=\cO \left( \eps^2\ln(\kzl) \right).
\end{equation}
In these equalities we have used the relations \eqref{rels} and \eqref{relkzleps}.


\end{document}